# Wideband Pulse Generation for Underwater Applications Using Parametric Array

Yasin Kumru, A. Sinan Taşdelen and Hayrettin Köymen

***Abstract*—We investigated wideband pulse generation for underwater acoustic applications using a parametric array. We fabricated a transducer consisting of a 3 mm thick 75 mm-by-75 mm square-shaped PZT ceramic plate, which is matched to water media at the radiating face and terminated by a very low impedance at the back. All measurements were made in a large test tank. We transmitted square-root amplitude modulated pulses centered around 855 kHz primary frequency. We showed that phase-sensitive generation of in-phase and out-of-phase bursts suitable for coded transmission using a parametric array is possible. We generated very short duration bursts, as short as half-cycle, at a 10-80 kHz difference frequency range. The definition of the bursts is excellent, e.g., with a normalized cross-correlation of 0.92 with an ideal 2-cycle square burst, for both in-phase and out-of-phase pulses.**

## I. Introduction

THE wideband pulses are essential for effective and reliable operations in various underwater acoustic applications [1], [2]. They provide better target detection and classification due to improved signal-to-noise ratio (SNR) and range resolution [3], [4]. They can mitigate interference from other acoustic sources, reflection, multipath propagation, reverberation, and diffraction [5]. They also provide high data rates, enabling real-time communication and data exchange [6], [7].

The problems raised by various underwater acoustic applications render wideband pulse generation challenging. Underwater acoustic applications using conventional arrays typically utilize narrowband pulses, particularly at low frequencies, tens of kHz range [8]. A straightforward approach to increase the bandwidth is to use short signals at the cost of SNR-related performance. However, employing this strategy is challenging due to the lack of an efficient wideband transducer operating at kHz frequencies [9]. The transmitted signals from conventional arrays also have broad beamwidths, causing energy loss and interference. Besides, low-frequency acoustic signals require larger physical arrays with sufficient element spacing to avoid spatial aliasing.

Specialized methods are employed in underwater applications to generate highly directive and almost side-lobe-free radiation at low frequencies. The advent of parametric arrays provides a means to obtain these features [10]-[13]. The parametric arrays exploit the inherent nonlinearity in the propagation in the medium, which occurs when the sound pressure amplitude is sufficiently high [14]. These arrays generate low-frequency narrow beams using a physically small transducer [15], [16]. They have been widely used underwater for acoustic characterization of materials [17], [18], communication purposes [11], [12], [19] sub-bottom profiling [20]-[22], range compensation [23], detection of buried targets [24]-[26], long-range ocean research [27], and sea surface scattering measurements [28]. They offer potential solutions to the problems of aero-acoustics [29]-[36] and biomedical imaging [37]-[39]. It was also stated in [17] that the parametric arrays enable wider bandwidth in underwater acoustic applications compared to their conventional counterparts.

Consequently, increasing the bandwidth as much as possible is critical in underwater acoustic applications. In this context, this study focuses on generating the wideband pulses, i.e., short-duration bursts, at kHz difference frequencies. To address this, we designed a wideband transducer with a center frequency, which is significantly higher than the difference frequency. We could thus maintain accurate phasing between the primary waves in a tradeoff with a lower secondary source level [7]. We used a parametric array in this study throughout the measurements performed in the test tank. The difficulties of establishing a parametric array in the confined space of the test tank are outlined. The experimental procedures and techniques used in this study to reduce the effects of these difficulties and process the measured data are described. We experimentally obtained well defined 2-cycle bursts at difference frequencies, although shorter signals were also observed.

Section II explains the difference frequency wave (DFW) generation using pulsed transmission. Section III presents the design methodology for the transducer used throughout the measurements. Section IV covers the measurement setup and procedure. Section V presents the experimental results. Finally, we discuss the results and present the conclusion in Section VI.

## II. Difference Frequency Wave Generation

The speed of sound in the medium is a fundamental constant in acoustic wave equation [40]. The speed of sound is a function of stiffness and the density of the medium. However, the amplitude of the propagating pressure wave locally modifies the density and consequently the speed of sound, when relatively higher-pressure levels are generated. Hence, acoustic wave propagation is inherently a nonlinear process.

The nonlinear interaction of two relatively high-frequency waves during their propagation through the same beam produces low-frequency, highly directive, and sidelobe-free radiation at the difference frequency [41]. These two high-

Electronic mail: yasin@ee.bilkent.edu.tr

frequency signals at frequencies $f_1$ and $f_2$ are primary signals, and the associated field is the primary acoustic field [7], [42]. A transducer transmits a linear combination of these primary signals. As the transmitted wave propagates underwater, components at new frequencies, such as a difference frequency component $(f_1 - f_2)$, sum frequency component, harmonics, and other higher-order terms arise due to the nonlinear interaction [42]. Among these new components, only the difference frequency component can travel to relatively larger ranges due to lower absorption [42]. The wave at the frequency of $f_d = f_1 - f_2$ is the DFW, and the associated field is the secondary acoustic field [42]. The remaining higher-frequency components vanish rapidly due to high absorption [42].

The radiated wave from the transducer is initially confined within the near-field region of the transducer [7]. In this region, the parametric conversion efficiency is high since the pressure wave amplitude maintains the highest levels throughout the region. The Rayleigh Distance, $R_F = S/\lambda_p$, where $S$ is the transducer surface area, and $\lambda_p$ is the wavelength at the average primary frequency, approximates the length of this region [43]. In the Fraunhofer Zone, the region beyond the near field, the wave spreads spherically, and wave intensity decays rapidly. Consequently, the parametric conversion efficiency decreases, resulting in a low-intensity DFW.

The propagating wave amplitude also decreases due to the acoustic absorption of the water medium. The absorption determines the effective parametric array length, also known as the absorption range [43]. The absorption range is expressed as $R_a = {a_p}^{-1}$, where $a_p$ is the absorption coefficient at the average primary frequency in nepers per unit length. The absorption coefficient can be calculated as in Eq. (1), where $f_T = 21.9 \times 10^{6-1520/(T+273)}$, T is the temperature in ℃, $f$ is the frequency in kHz, and $S_a$ is the salinity in ‰ [43].

$$\alpha\left(\frac{dB}{kyd}\right) = \left(1.86 \times 10^{-2}\frac{S_a\, f_T\, f^2}{{f_T}^2 + f^2}\right) + \left(2.68 \times 10^{-2}\frac{f^2}{f_T}\right) \quad (1)$$

The shock distance is the other crucial distance for the DFW generation [7]. Due to nonlinear effects on the primary waves, the sound speed becomes a function of the primary wave amplitude. The high-amplitude wave crests travel faster than the low-amplitude wave troughs. As a result, the sine wave starts to distort and becomes a saw-tooth wave after the primary wave travels a certain distance. This distance is the "shock distance," $R_s = (c_0\lambda_p)/(4\beta u)$ [7], where $c_0$ is the sound speed in the medium, $\beta$ is the nonlinearity parameter, and $u$ is the particle velocity in the primary wave. The distortion shifts some of the energy in the primary wave to the harmonic components, resulting in energy loss called excess attenuation [44].

### A. Alternative way for difference frequency wave generation

It is possible to operate the transducer using either two primary frequencies or a pulse consisting of a modulated carrier signal [18]. In the latter case, the nonlinear effects demodulate the envelope of the carrier signal. This process, known as the self-demodulation, produces DFW [44], [45]. When a pulse consisting of a carrier signal modulated with an envelope function, $E(t)$, is transmitted, the pressure of the DFW at the secondary acoustic field is [13]:

$$p_d(t) = \frac{\beta p_0^2 S_0}{16\pi\rho_0 c_0^4 r\alpha_0}\frac{d^2}{dt^2}E^2(t) \quad (2)$$

Equation (2) is known as Berktay's far-field solution for the DFW pressure. It states that the pressure at the difference frequency depends on the second time derivative of the square of the envelope of the transmitted amplitude-modulated carrier signal. The envelope function, $E(t)$, has a maximum value of unity. In Eq. (2); $p_o$, $S_0$, $\rho_0$, $r$, and $\alpha_0$ stand for the pressure amplitude of the primary signal, cross-sectional area of the sound beam, medium density, propagation distance, and absorption coefficient at the primary frequency, respectively.

### B. Method

We used the self-demodulation process to generate very wideband pulses at the difference frequency. Short pulses offer wider bandwidth and higher resolution at the cost of reduced available energy and related performance. Coded signals are a better alternative to short pulses as they increase the transmitted signal energy at a given peak transmitted pressure without sacrificing available bandwidth [46].

In this regard, we constructed the transmitted signals to generate the desired wideband pulses at the difference frequency for symbols representing "+1" and "-1" bits of the binary phase shift keying modulated coded signal. We used the relation between the transmitted signal envelope and DFW as stated in Eq. (2). By double integrating both sides of Eq. (2), we obtained the necessary envelope function of the transmitted signal, which allowed us to generate the desired pulse at the difference frequency. We constructed the envelope functions such that the pulses at the difference frequency representing "+1" and "-1" symbols of a coded signal are 180° out of phase.

### C. Transmitted pulsed waveform

We defined the envelope functions of the transmitted signals necessary to generate the pulses for symbols representing "+1" and "-1" bits at the difference frequency. These pulses are denoted as $p_d(t)$ and $p'_d(t)$, respectively. We used the square-root amplitude modulation technique (SQRAM) to generate the transmitted signals since it is commonly used in literature to reduce the distortion in the DFW [44]. We used the envelope functions expressed in Eqs. (3) and (4) to produce 180° phase difference at $f_d$. Here, $\omega_d$ is the angular frequency at $f_d$. Note that, these envelope functions exhibit 90° phase difference.

$$E_s(t) = \sqrt{1 - sin(\omega_d t)} = \sqrt{2}\left|cos(\frac{\omega_d t}{2} + \frac{\pi}{4})\right| \quad (3)$$

$$E_{s'}(t) = \sqrt{1 + sin(\omega_d t)} = \sqrt{2}\left|cos(\frac{\omega_d t}{2} - \frac{\pi}{4})\right| \quad (4)$$

We obtained the transmitted signal by multiplying the envelope functions with a sinusoidal carrier signal. The resulting amplitude-modulated carrier signal, $s(t)$, with an envelope function, $E_s(t)$, is given in Eq. (5). Here, $\omega_c$ is the angular frequency of the carrier signal at 855 kHz, which is the resonance frequency of the transmitter, and $p(t)$ is the rectangular pulse restricting $s(t)$ to the interval $0 \leq t \leq t_d$, where $t_d$ is the time duration of the DFW to be generated.

$$s(t) = E_s(t)\, sin\,(\omega_c t)\, p(t) \quad (5)$$

We also constructed a transmitted signal, $s'(t)$, to generate a pulse for symbol representing "-1" bit at the difference frequency, ${p'}_d(t)$, which is expected to be 180° out of phase compared to $p_d(t)$. Similarly, we obtained the transmitted signal exactly as defined in Eq. (5), but with the envelope function replaced by $E_{s'}(t)$.

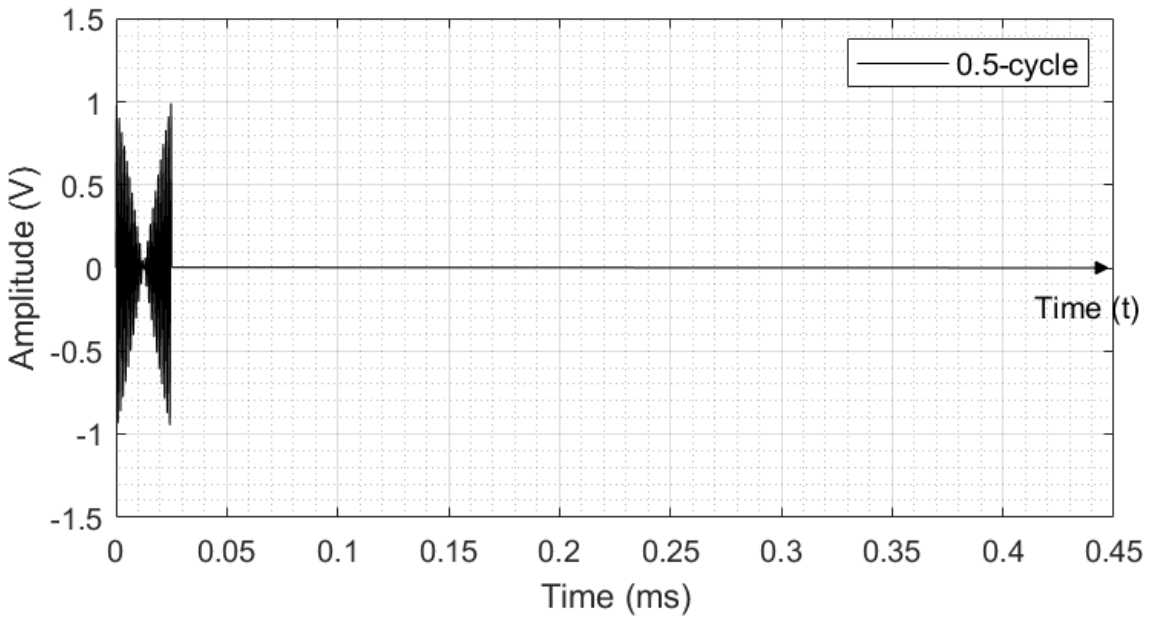

(a)

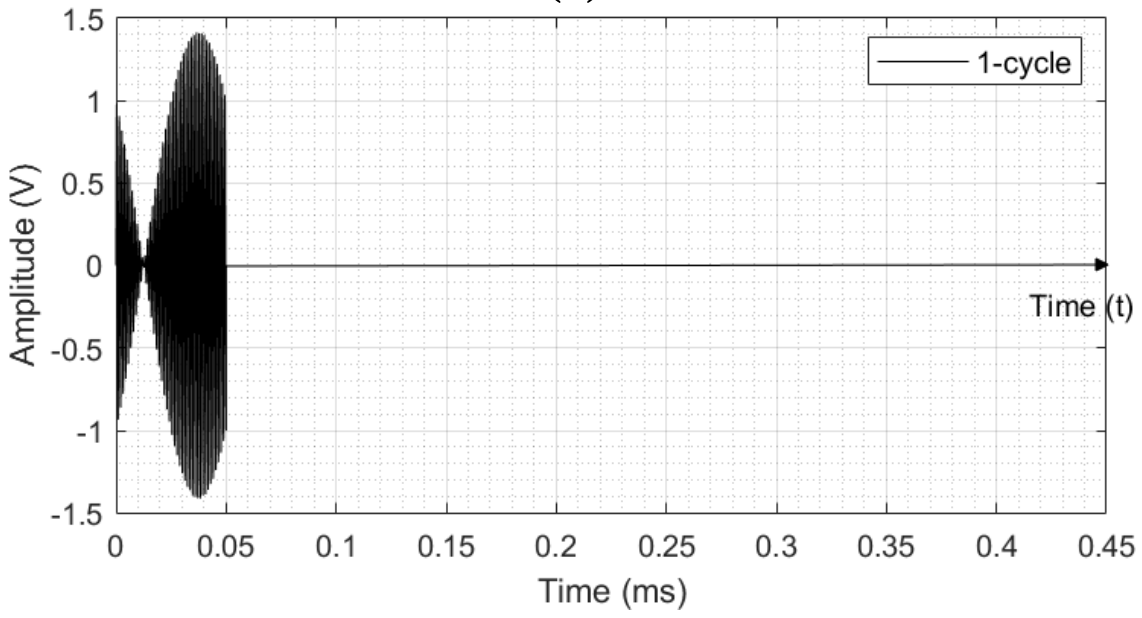

(b)

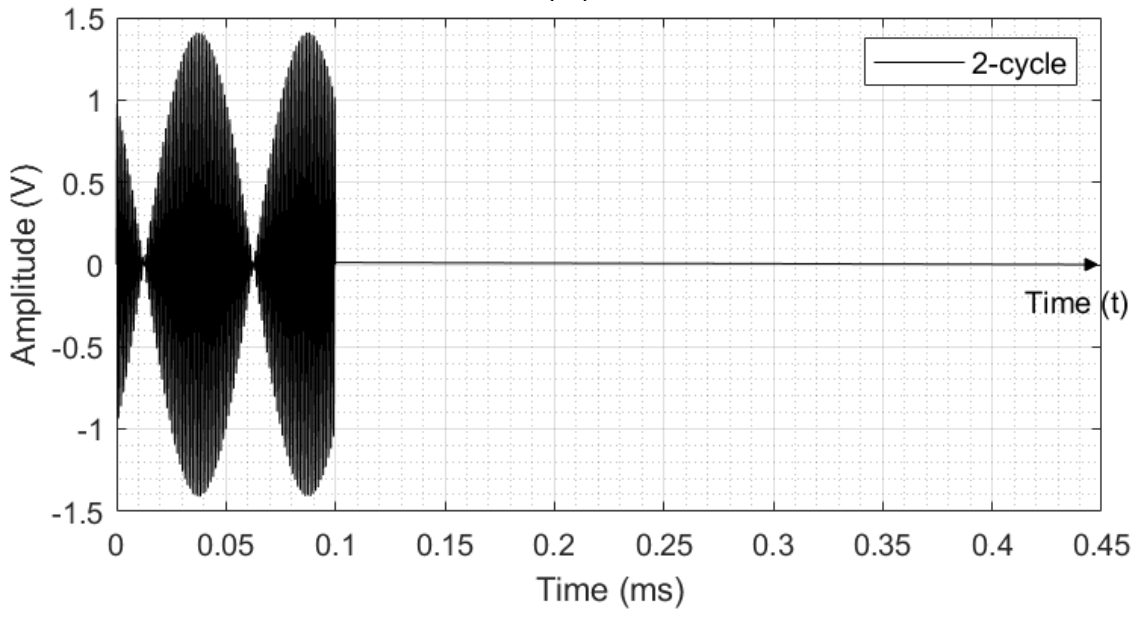

(c)

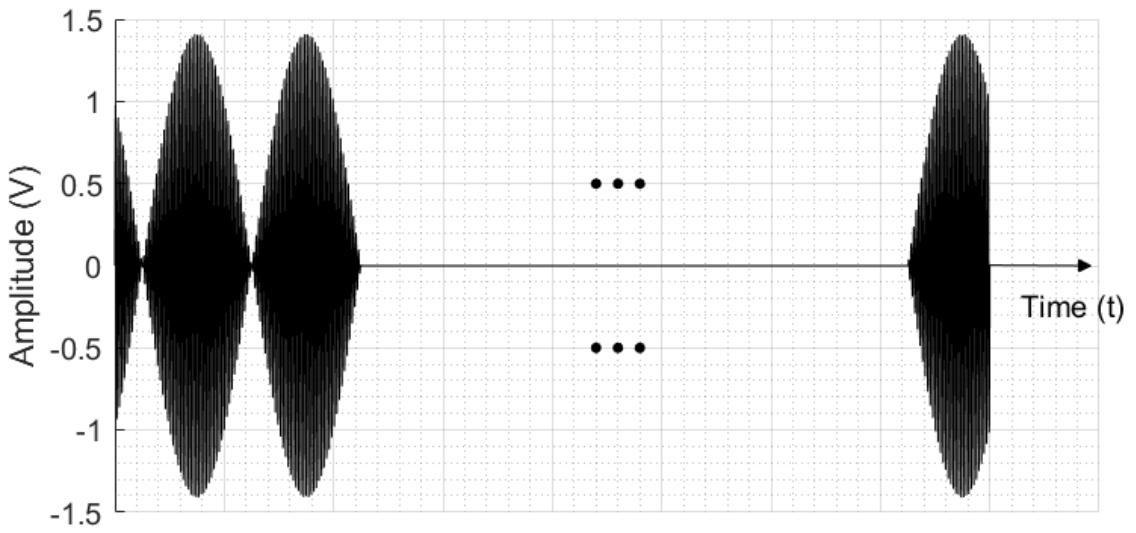

(d)

Fig. 1. The transmitted amplitude-modulated pulsed waveforms, $s(t)$, to generate (a) 0.5-cycle (b) 1-cycle (c) 2-cycle pulses $(d)$ pulses with more cycles at 20 kHz difference frequency. "V" represents for volts. These signals are transmitted through the medium to produce difference frequency waves.

Figures 1(a)-1(c) show the transmitted signals, $s(t)$, used for generating 0.5, 1, and 2-cycle pulses, $p_d(t)$, at a 20 kHz difference frequency, respectively. Figure 1(d) depicts the transmitted signal waveform for pulse generation with more cycles at the difference frequency. The envelope signal exhibits a $\pi/4$ phase difference as shown in Eq. (3), and it is visible in Fig. 1. All signals except for the transmitted signal required for the 0.5-cycle pulse generation at 20 kHz reach their peak values. However, the transmitted signal required for the 0.5-cycle pulse generation, shown in Fig. 1(a), fails to attain its peak value, which indicates the need for a different modulation technique.

Figures 2(a)-2(c) show the transmitted signals, $s'(t)$, for generating 0.5, 1, and 2-cycle pulses at 20 kHz, ${p'}_d(t)$ (the out-of-phase symbols), respectively. For pulse generation with more cycles at the difference frequency, the transmitted signal generation is shown in Fig. 2(d). The $-\pi/4$ phase difference in the envelope signal, as shown in Eq. (4), is evident in Fig. 2. Comparing the signals in Figs. 1 and 2 reveals a $\pi/2$ phase difference. According to Eq. (2), this phase difference in the transmitted signals is expected to result in 180° phase difference in the secondary acoustic field after nonlinear interaction.

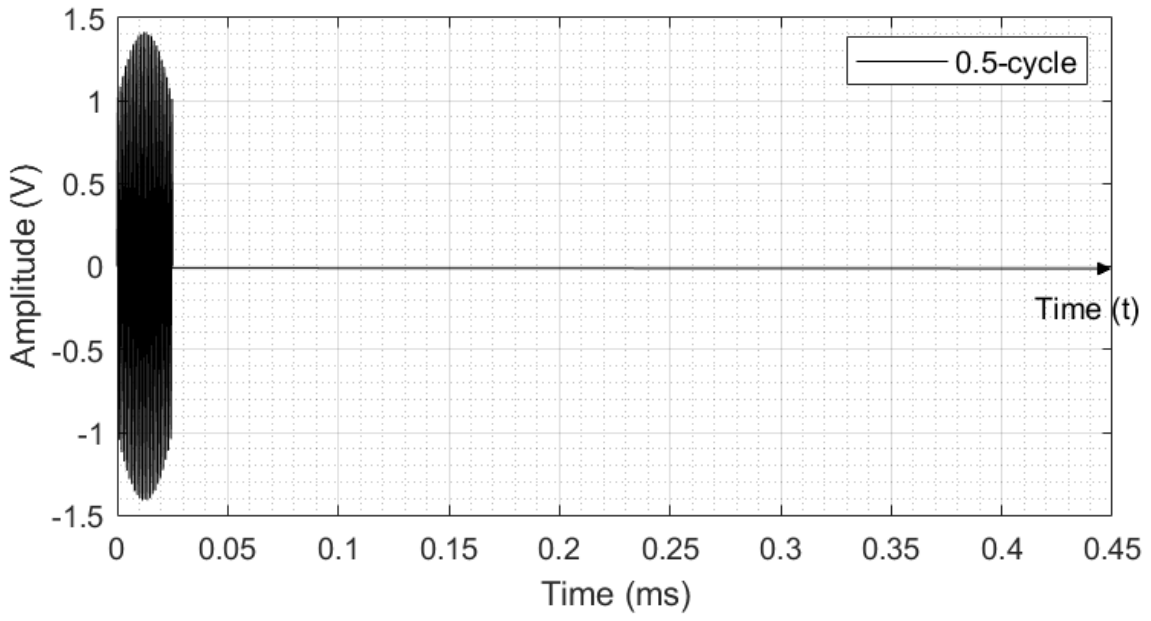

(a)

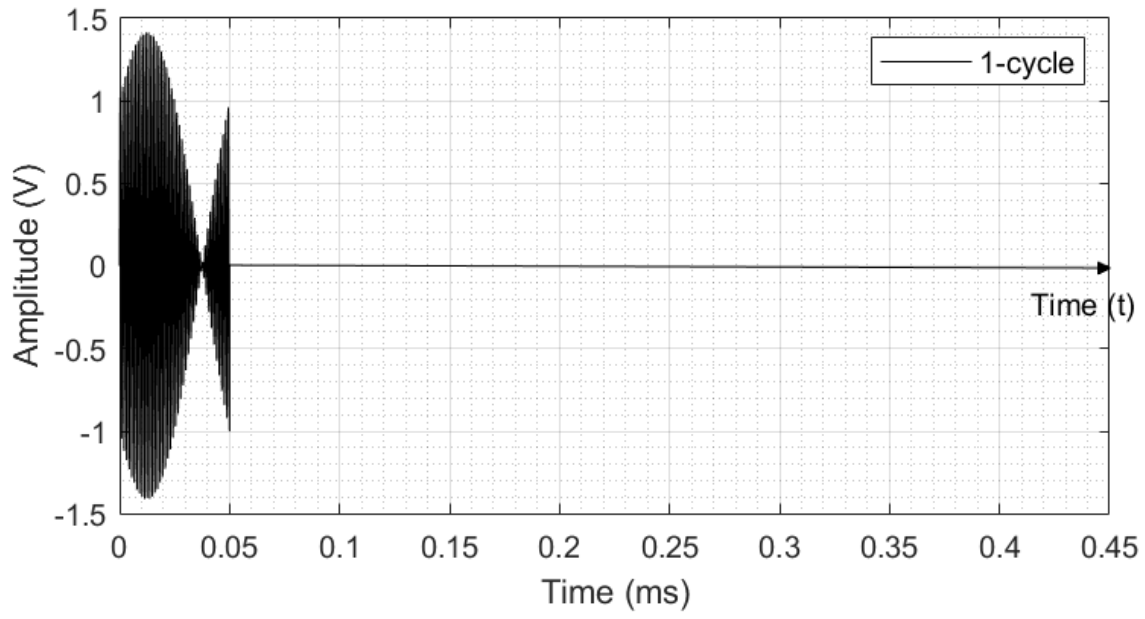

(b)

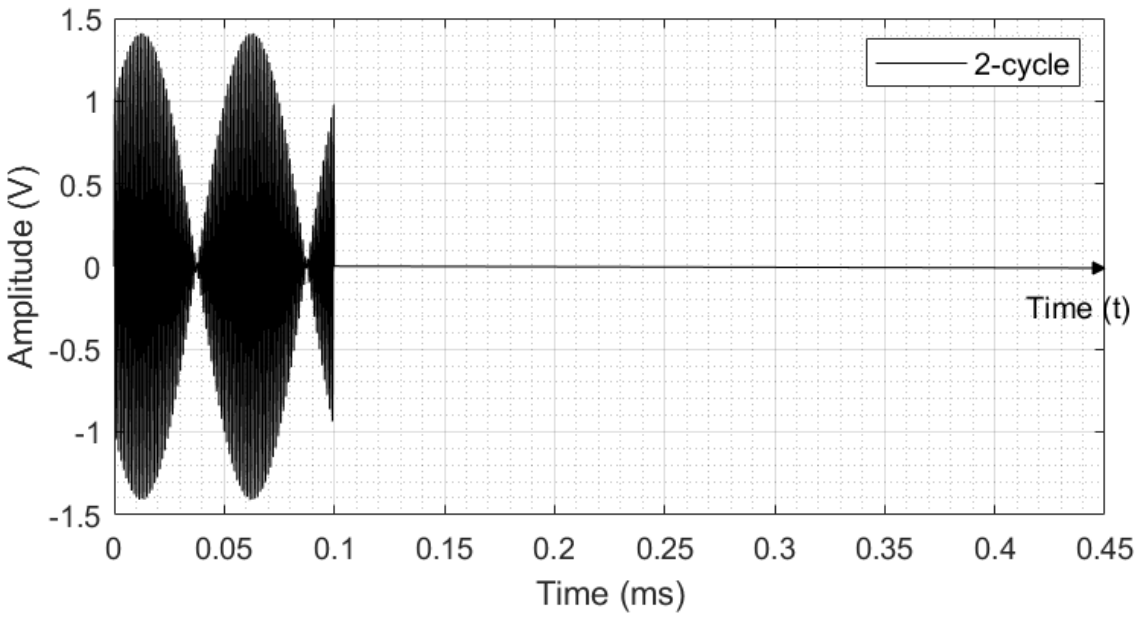

(c)

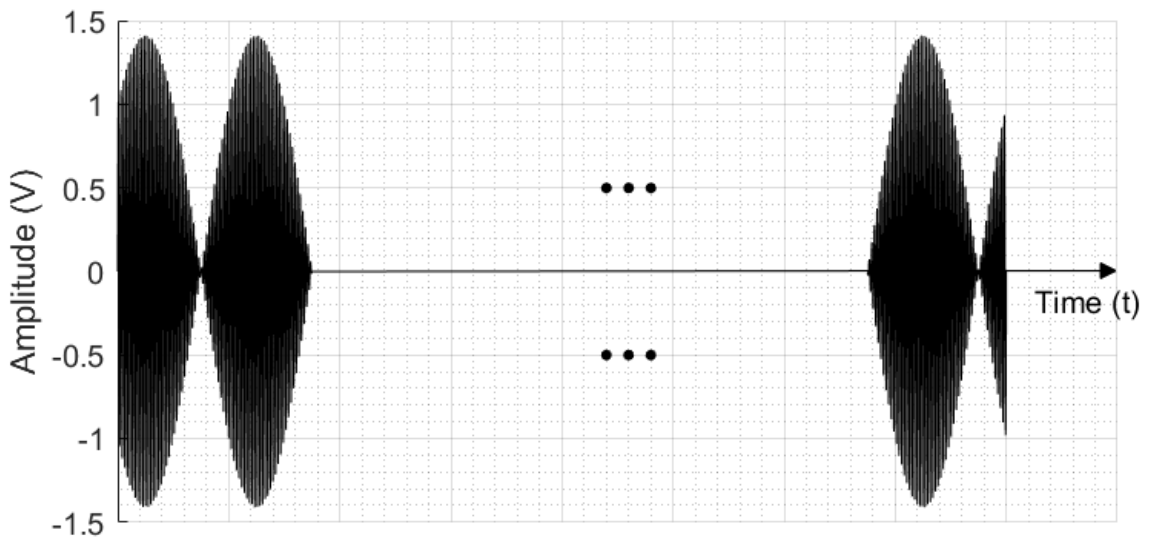

(d)

Fig. 2. The transmitted amplitude-modulated pulsed waveforms, $s'(t)$, to generate (a) 0.5-cycle (b) 1-cycle (c) 2-cycle pulses (d) pulses with more cycles at 20 kHz. These signals are transmitted through the medium to produce DFWs.

## III. The Transducer For Parametric Array

To experimentally validate the wideband pulse generation in the secondary acoustic field, we first designed and fabricated a piezoelectric transducer operating at 855 kHz center frequency as a source. The transducer and its cross section are shown in Figs. 3(a) and 3(b), respectively. The equivalent electrical circuit of this transducer is also shown in Fig. 4 [47]. This equivalent circuit comprises a piezoelectric plate (or ceramic), backing material, matching layer, and radiation medium.

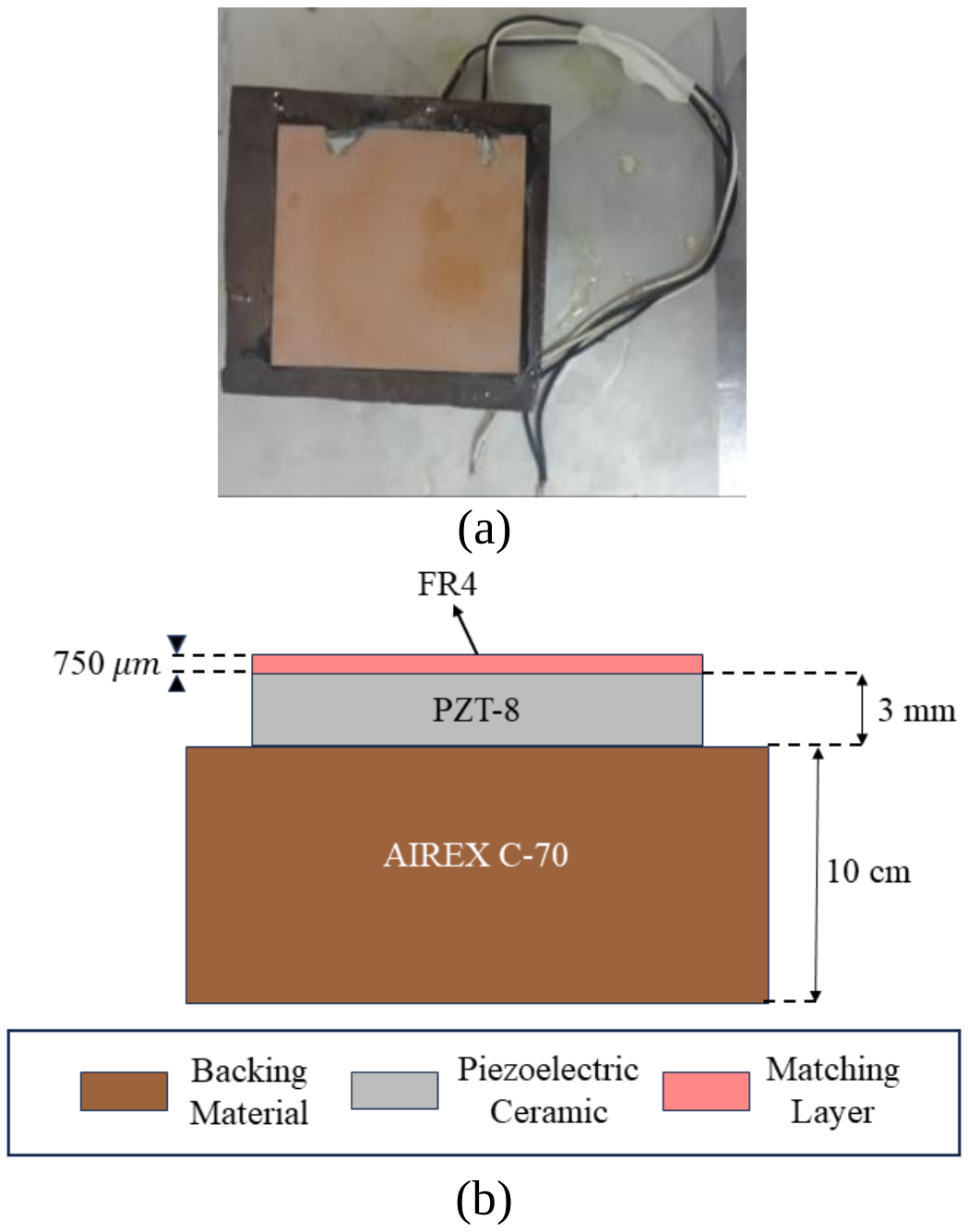

Fig. 3. (a) Image of the transducer. (b) Cross-section of the transducer.

We used a commercially available Lead-Zirconate-Titanate (PZT-8) ceramic as the active component of the transducer. It was 75 mm-by-75 mm square-shaped ceramic plate with a thickness of 3 mm.

The transducers have a matching layer on the transducer's radiating surface side, as shown in Fig. 4. This layer matches the load impedance to the transducer [47]. We used FR4 (fibre reinforced polymer-grade 4) as a matching layer. We bonded FR4 to the ceramic using epoxy resin (Huntsman Corporation, Model Araldite 2011). We used an epoxy-curing oven to heat the mixture of the epoxy resin and its hardener. We completed the curing process in 4 hours at a temperature of 120 ℃. The optimal operating thickness for the matching layer is one-fourth of the wavelength in the piezoelectric ceramic at 855 kHz. The optimal thickness corresponds to 750 $\mu$m, while the thickness of the matching layer is 1.1 mm. We refined the matching layer's thickness through lapping until the optimal thickness was reached.

On the transducer's back surface, we used a polymer foam, AIREX C-70, as a backing material. We bonded the ceramic to the backing material using the same bonding and curing processes we used for the matching layer.

Lastly, it is also crucial to consider the radiation medium as an integral component of the transduction [47]. It manifests itself as an impedance called radiation impedance, $Z_R(ka)$, in the transduction mechanism as shown in Fig. 4. The radiation impedance depends on the wavenumber, $k$, and the radiation aperture. The real part of the radiation impedance, the radiation resistance, $R_R(ka)$, determines the radiated acoustic power into the medium [47].

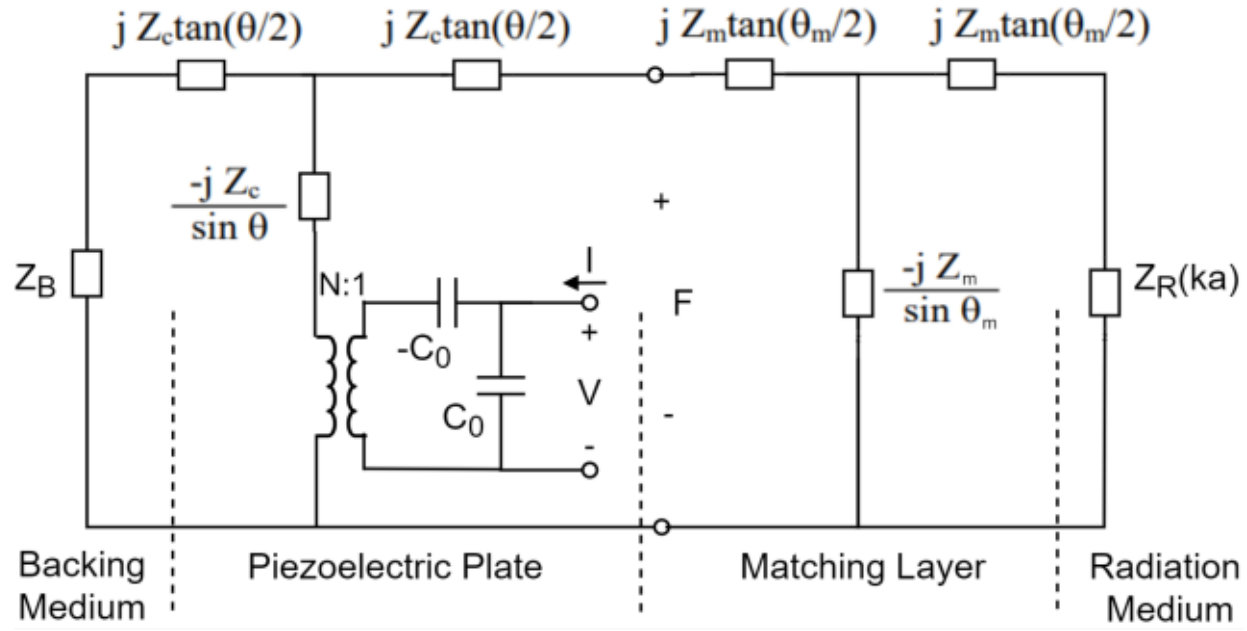

Fig. 4. The equivalent circuit of the transducer. This circuit consists of backing medium, piezoelectric plate, matching layer, and radiation medium.

We measured the admittance of the transducer, shown in Fig. 3(a), in freshwater. Figure 5 shows the real (conductance) and imaginary parts (susceptance) of the admittance. The resonance frequency of the transducer is measured as 855 kHz.

There are two observations on the variation of impedance with respect to frequency: i) the susceptance is predominantly due to the clamp capacitance ($C_0$) at the electrical port, approximately 17 nF, and ii) the conductance promises a very wideband operation, about 400 kHz[1], with high efficiency, due to a very efficient impedance transformation and matching at the acoustic port [48].

Making use of this wideband performance available at the acoustic port despite very large clamp capacitance requires very sophisticated matching circuits at the electrical port. Simple tuning using a parallel or series inductor yields an efficient but a rather narrow band operation. However, for only transmission purposes as in the case of a parametric array, the transducer can be driven using a "voltage source", thus eliminating the effect of shunt reactance at the electrical port. Practically this corresponds to driving the transducer using a power amplifier where the clamp capacitance reactance is significantly larger than amplifier output impedance. This means that most of the available power at the driver is wasted to take care of the clamp reactance. The clamp reactance at the center frequency is about 11Ω in this transducer and a driver with much lower output impedance is not available. Therefore, we made a compromise in our measurements and tuned the transducer using a series 2.1 μH inductor at 855kHz, the center frequency. We used this transducer throughout the measurements and measured the transmitted acoustic signal in the medium in all cases.

[1] The presence of very high clamp capacitance complicates the direct impedance measurement in the water tank above 1 MHz. The inductance of leads (although quite short, about 75 cm) in the measurement setup are significant and resonates with the clamp capacitance at just above 1 MHz. Hence lead compensation at the impedance analyzer is not effective. This measurement artifact manifests itself in the impedance plot as a sustained conductance level above 1 MHz, whereas there should be a fall off, limiting the upper 50% level to about 1.05-1.1 MHz.

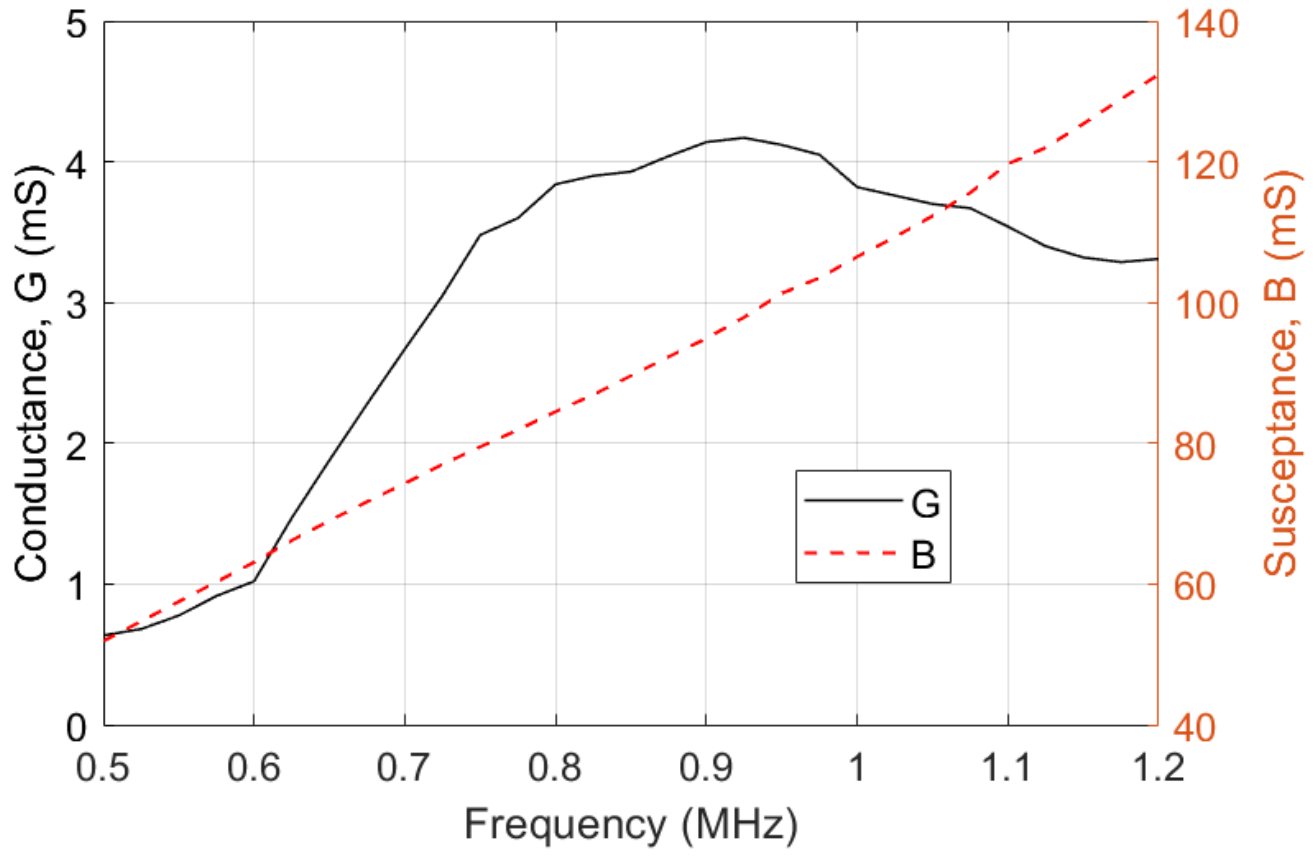

Fig. 5. Measured conductance (G) and susceptance (B) of the transducer. The transducer has a sufficient bandwidth around 855 kHz resonance frequency.

## IV. Measurement Set-Up And Procedure

### A. *Measurements and equipment*

We carried out measurements in a 3.5 m long × 2 m wide test tank filled with freshwater to a depth of about 1.8 m. We measured the primary pressure wave, DFW, and transducer beamwidth. Figure 6(a) shows the positions of the transducer and two hydrophones in the tank, and Fig. 6(b) shows the block diagram of the measurements.

We deployed the transducer at one end of the tank at a depth of about 1 m. It is driven using a signal generator (Stanford Research Systems, Model DS345) and a power amplifier (Krohn-Hite Corporation, Model 7500). We created waveforms using an arbitrary waveform composer software installed on a personal computer and downloaded the waveforms to the DS345 [49]. The function generator output was fed to the power amplifier. We adjusted the power amplifier gain to prevent saturation during the transmission.

We measured the primary wave using a needle hydrophone (Precision Acoustics, 2.0 mm Needle Hydrophone, Model NH2000) [50] to avoid any distortion in the primary field. We deployed this hydrophone 1 m away from the transducer. It is designed to be used in conjunction with a submersible preamplifier and a DC coupler with a power supply. The hydrophone output was directly connected to the oscilloscope (Agilent Technologies, Model DSO1002A). We recorded the measured primary wave displayed on the oscilloscope screen for subsequent use.

We measured the pressure of the DFW using a D/140/H hydrophone (Neptune Sonar Ltd., Model D/140/H) [51] combined with a submersible preamplifier. We deployed this hydrophone 2 m away from the transducer. The preamplifier of this hydrophone was powered from an external 24 Volt DC power supply (Agilent Technologies, Model E3616A). The output of this hydrophone was connected to the low noise amplifier (LNA, Stanford Research Systems, Model SR560). Since the propagation distance is limited due to the finite tank size, we had to measure the DFW at 2 m range, where the primary field is very high. Consequently, we had to filter out the contribution of the primary frequency waves in the measurements. We used LNA as a bandpass filter with 6 dB roll-off and cutoff frequencies of 1 kHz and 300 kHz to filter out this contribution.

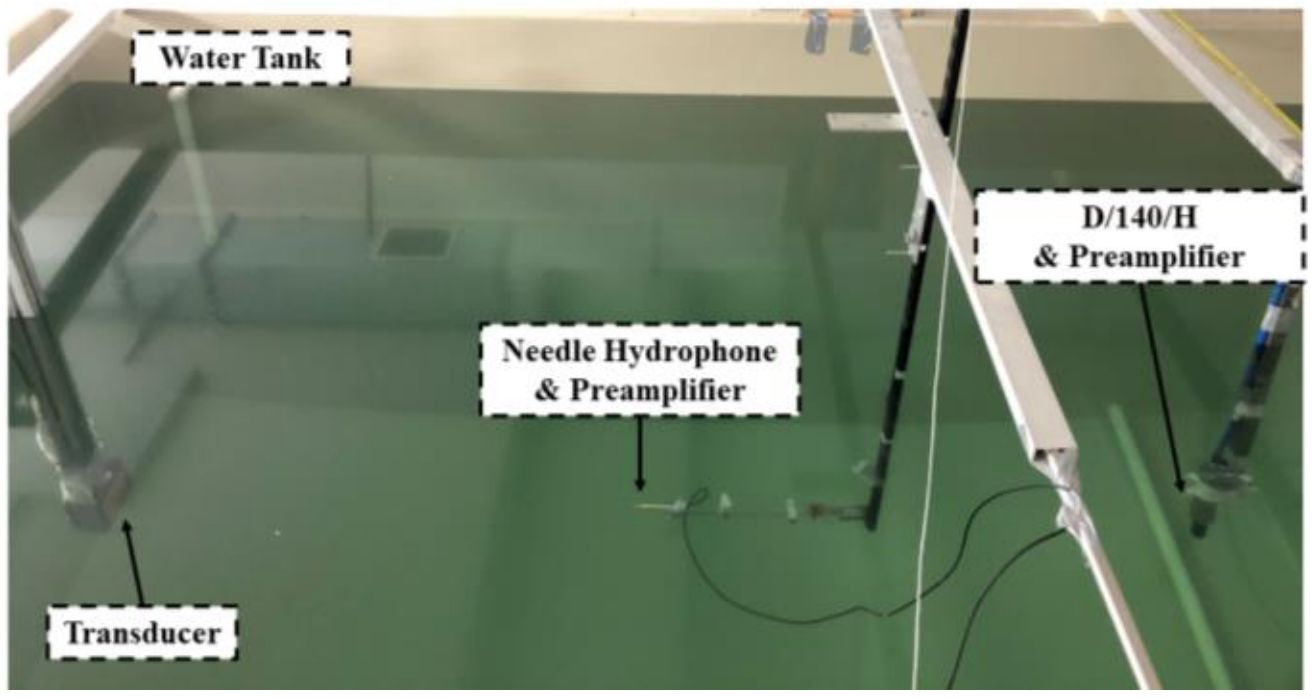

(a)

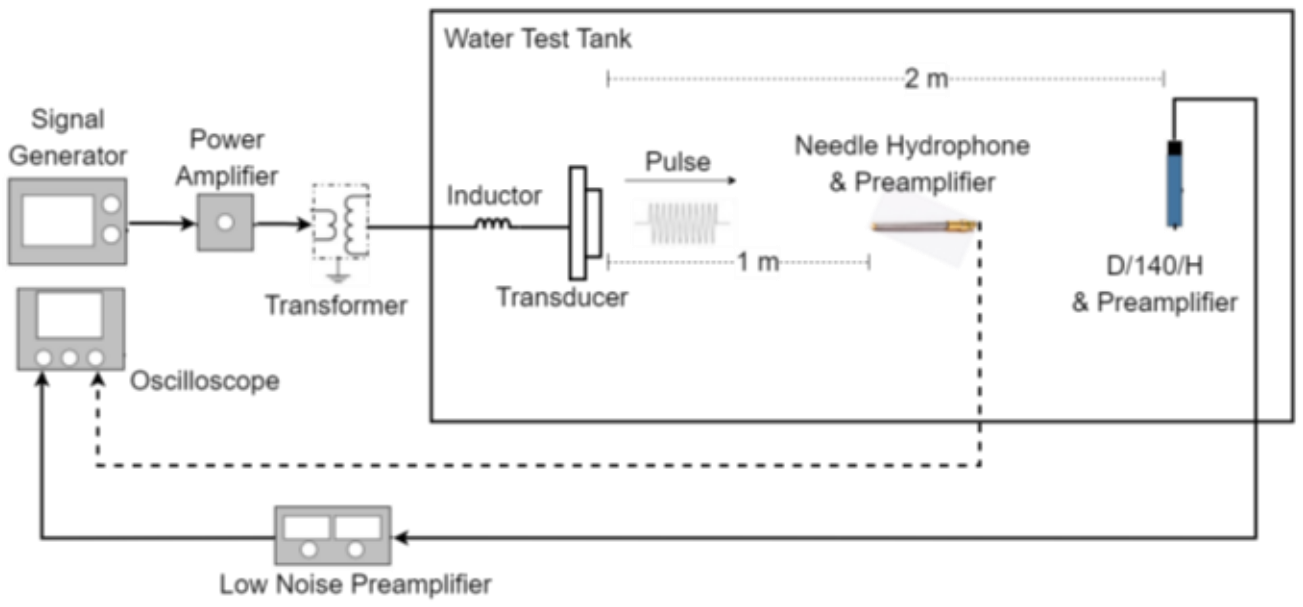

(b)

Fig. 6. (a) The positioning of the transducer, needle and D/140/H hydrophones in the test tank. (b) Block diagram for the measurements. The connection of the needle hydrophone is shown with a dashed line since it is removed from the water after the primary wave measurement was performed.

First, we measured the pressure wave at the primary frequency using a needle hydrophone in every measurement. Subsequently, we removed the needle hydrophone from the water and measured the pressure wave at the difference frequency by the D/140/H hydrophone. We further post-processed the data measured by D/140/H hydrophone to filter out the undesired frequency components, which are still present even after using LNA as a bandpass filter. For this purpose, we implemented a software low-pass filter. The order of the low-pass filter is 28. The cutoff frequency was set to twice the difference frequency. We also implemented a second-order high-pass filter with a 1 kHz cutoff frequency to remove the DC offset. We implemented Butterworth filters in MATLAB for these filtering operations.

### B. *Critical distances affecting secondary wave measurements*

The absorption range in water, $R_a$, at 855 kHz is 38 m, which is very long compared to any other distance affecting the secondary wave generation in this work. In this respect, it is well justified to assume that the primary wave is not attenuated at all during the nonlinear interaction process.

The shock distance, $R_s$, for the transmit power levels delineated in the following section (Sec. V) is about 14.5 m at 855 kHz. The Rayleigh distance for the transmitting transducer is 3.2 m, which is significantly shorter than the shock distance.

The measurement tank dimensions limit the measurement distance to 2 m, which is shorter than the Rayleigh distance. Hence, the potential for secondary wave generation in the parametric array is not fully exploited at this measurement distance. Therefore, the measured secondary field source levels are significantly lower than what it is in the far-field measurements and the beamwidth is expected to be narrower than the measured secondary field beamwidth.

## V. Measurement Results

### A. Primary acoustic field measurements

We transmitted amplitude-modulated signals given in Figs. 1(c) and 1(d) to generate 2, 4, and 8-cycle pulses representing the "+1" symbol at 20 kHz difference frequency. Then, we measured the pressure in the primary acoustic field, denoted as $p_N(t)$, using the needle hydrophone. Figures 7(a)-7(c) show these measured pressure signals for 2, 4, and 8-cycle pulse cases, respectively.

In all measurements, the pressure wave reached the hydrophone approximately at 0.63 ms, which is consistent with the expectations based on the sound speed and distance between the transducer and the needle hydrophone. We measured approximately 20 kPa peak pressure at 1 m from the transducer for all transmissions.

We observed that the envelope of the measured pressure signals shown in Fig. 7 remains preserved compared to those shown in Fig. 1. This preservation indicates that the transmitted signal maintains its original shape and characteristics upon transmission through the transducer and propagation in water. Additionally, the signals reflected from the surface, bottom, and sidewalls of the tank and received by the needle hydrophone do not cause any distortion in the direct path signal. These spurious signals arrive at the measurement point at different delays due to the transducer and hydrophone placement. As a result, the integrity of the transmitted signal is not significantly affected by the propagation conditions in the test tank, which ensures reliable results in the secondary acoustic field.

The duration of the measured pressure signals increases due to the transducer transient response, as illustrated in dashed ellipse in Fig. 7(a) for the 2-cycle case. The similar effect of this response is also evident in Figs. 7(b) and 7(c). The signal originating from the transducer transient response starts immediately after the primary pressure signal ends. The amplitude of this signal is significantly lower than that of the primary pressure signal shown in the dotted rectangle in Fig. 7(a). Given that nonlinear interaction occurs at high signal amplitudes, the contribution of this signal to the nonlinear interaction turns out to be negligible and has no effect on the secondary acoustic field.

Similarly, we conducted another set of measurements involving the transmission of amplitude-modulated signals shown in Figs. 2(c) and 2(d). We transmitted these signals to generate 2, 4, and 8-cycle pulses representing the "-1" symbol at 20 kHz. We again measured the pressure in the primary acoustic field, denoted as $p'_N(t)$, using the needle hydrophone for all transmissions as shown in Fig. 8. We observed similar results in Fig. 8 compared to those in Fig. 7, as expected.

We also observed that transmissions aimed at generating DFW with higher difference frequencies exhibit more distortion in the primary acoustic field. This deterioration is clearly visible in Fig. 9. Figure 9 shows the measured pressure waves in the primary acoustic field in response to the transmissions aimed at generating 4-cycle pulses representing the "+1" symbol at 10, 20, 40, and 80 kHz difference frequencies, respectively.

For the 20 kHz signal, shown in Fig. 9(b), the carrier wave between 0.64 ms and 0.69 ms (the first cycle of the amplitude modulated waveform) has a lower amplitude compared to the remaining signal part beyond 0.69 ms. This observation is also evident in Fig. 7, in which 20 kHz DFWs for "+1" symbol are also aimed to be generated. Similarly, the 40 kHz signal showed a decrease in amplitude for the first two cycles compared to the last two. The 80 kHz signal demonstrates a noticeable reduction in amplitude across all cycles. This decrease in amplitude at 20, 40, and 80 kHz is attributed to the narrowing of the transducer bandwidth caused by the use of a 2.1 μH inductor for electrical tuning. As a result, components of the signals, which fall outside this narrower band, were affected. This manifested itself as a reduction in pressure amplitude. In contrast, the 10 kHz signal fits best within the transducer bandwidth due to its narrower bandwidth than other signals. Note that, the duration of the measured pressure wave reduces to half as the frequency doubles. Thus, the 10 kHz signal has the largest duration and the narrower bandwidth. Consequently, it remained unaffected and maintained consistent amplitude across all four cycles.

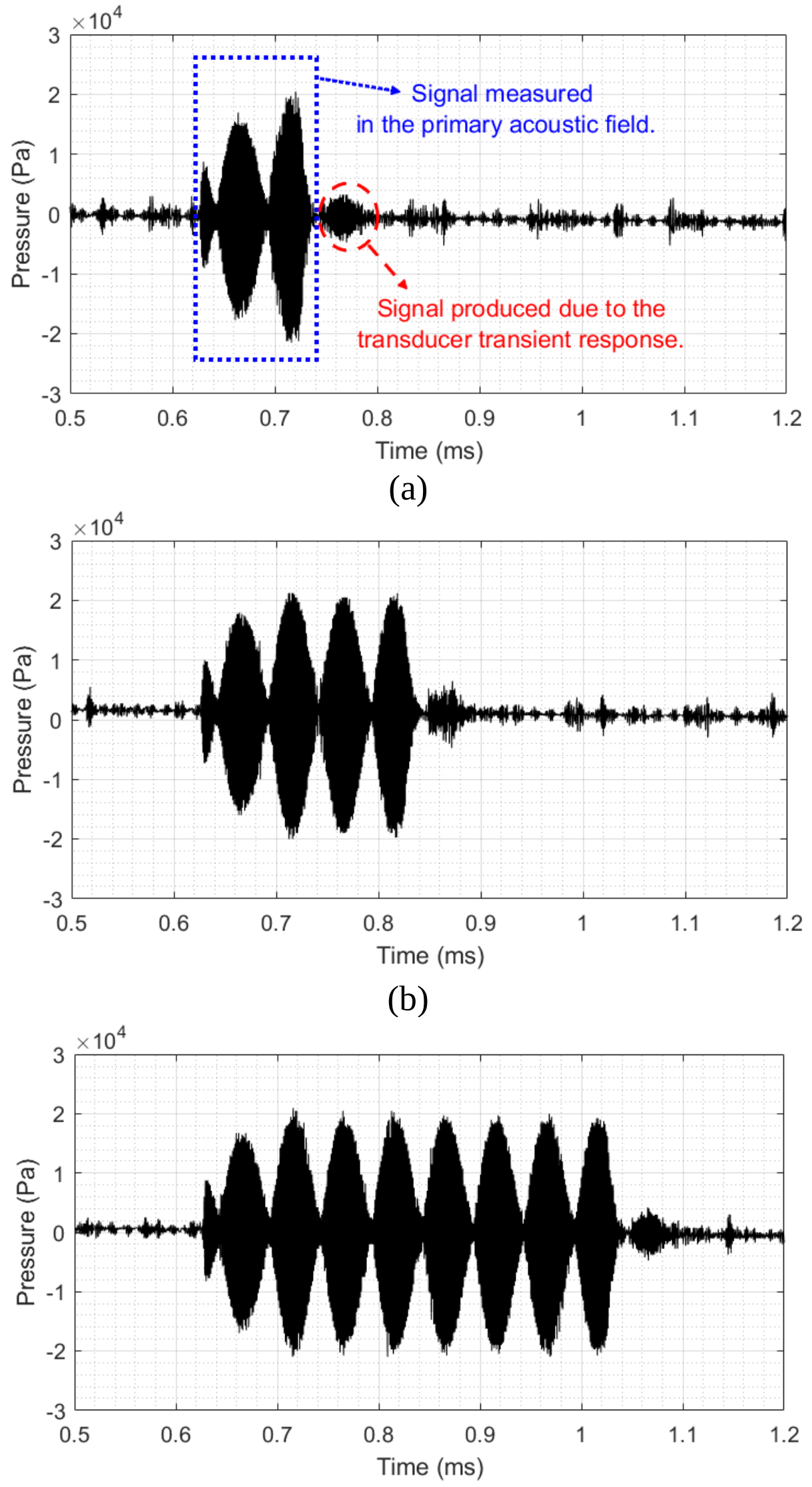

(a)
(b)
(c)

Fig. 7. The pressure waves, $p_N(t)$, sensed by the needle hydrophone at 1 $m$ away from the transducer, resulting from the transmission of $s(t)$. The subscript "N" represents the needle hydrophone. These waves are propagating waves in the primary acoustic field, which will result in (a) 2-cycle (b) 4-cycle (c) 8-cycle pulses at 20 kHz difference frequency in the secondary acoustic field. The ellipse shows the effect of the transducer transient response. The rectangular box shows the pressure wave measured in the primary acoustic field.

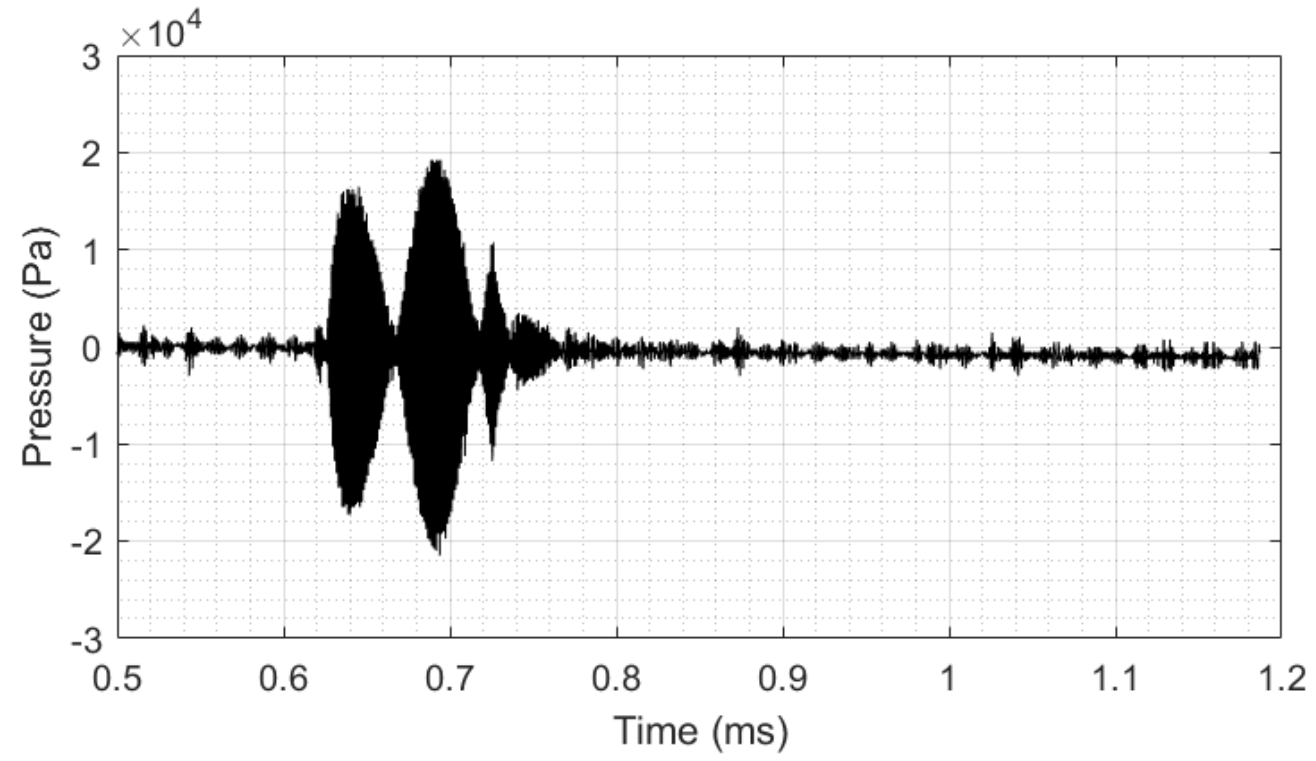

(a)

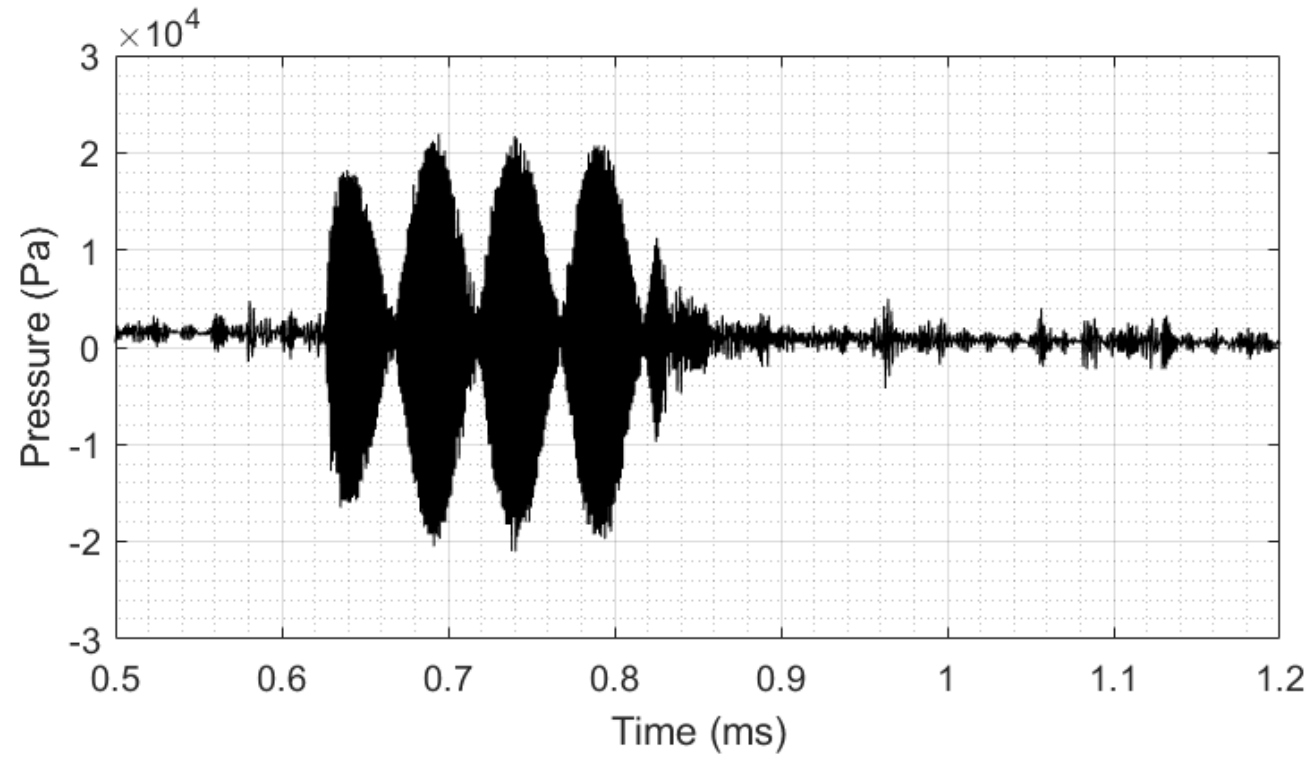

(b)

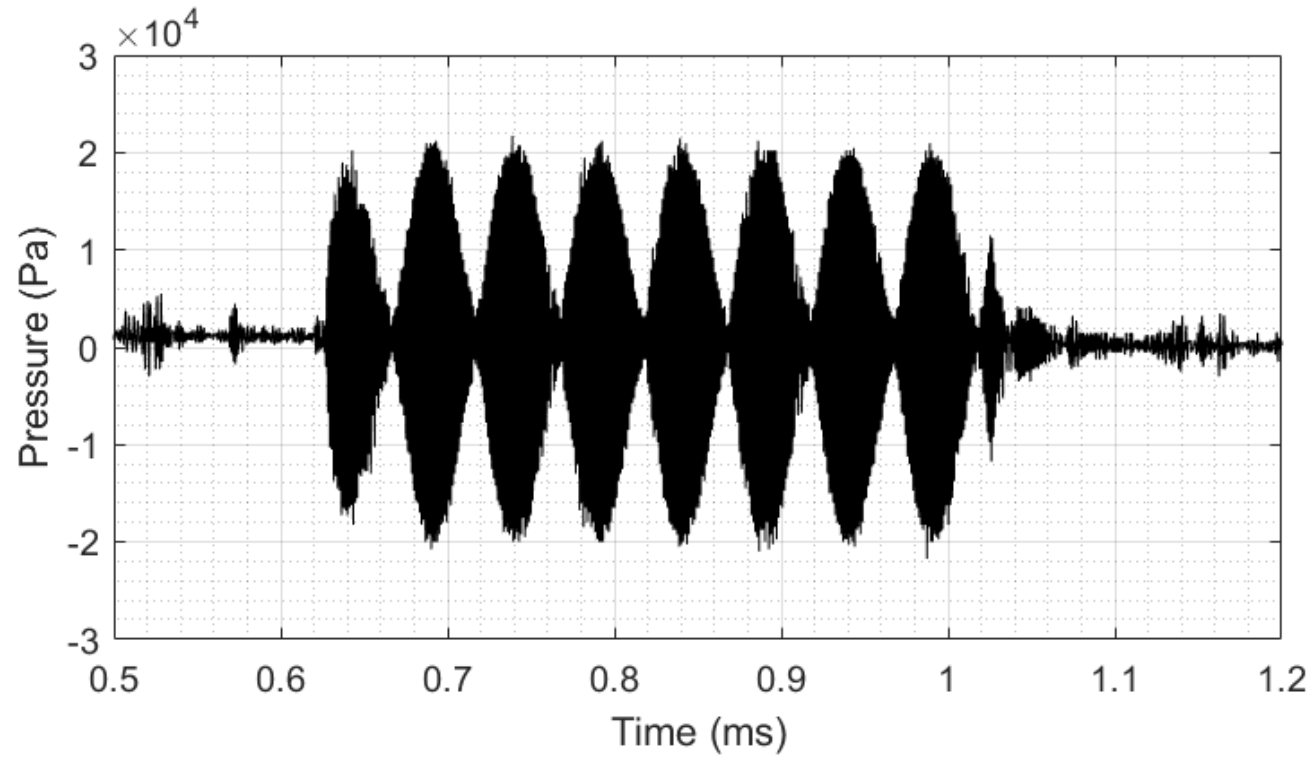

(c)

Fig. 8. The pressure waves, $p'_N(t)$, sensed by the needle hydrophone at $1\ m$ away from the transducer, resulting from the transmission of $s'(t)$. The subscript "N" represents the needle hydrophone. These waves are propagating waves in the primary acoustic field, which will result in (a) 2-cycle (b) 4-cycle (c) 8-cycle pulses at 20 kHz difference frequency in the secondary acoustic field after nonlinear interaction.

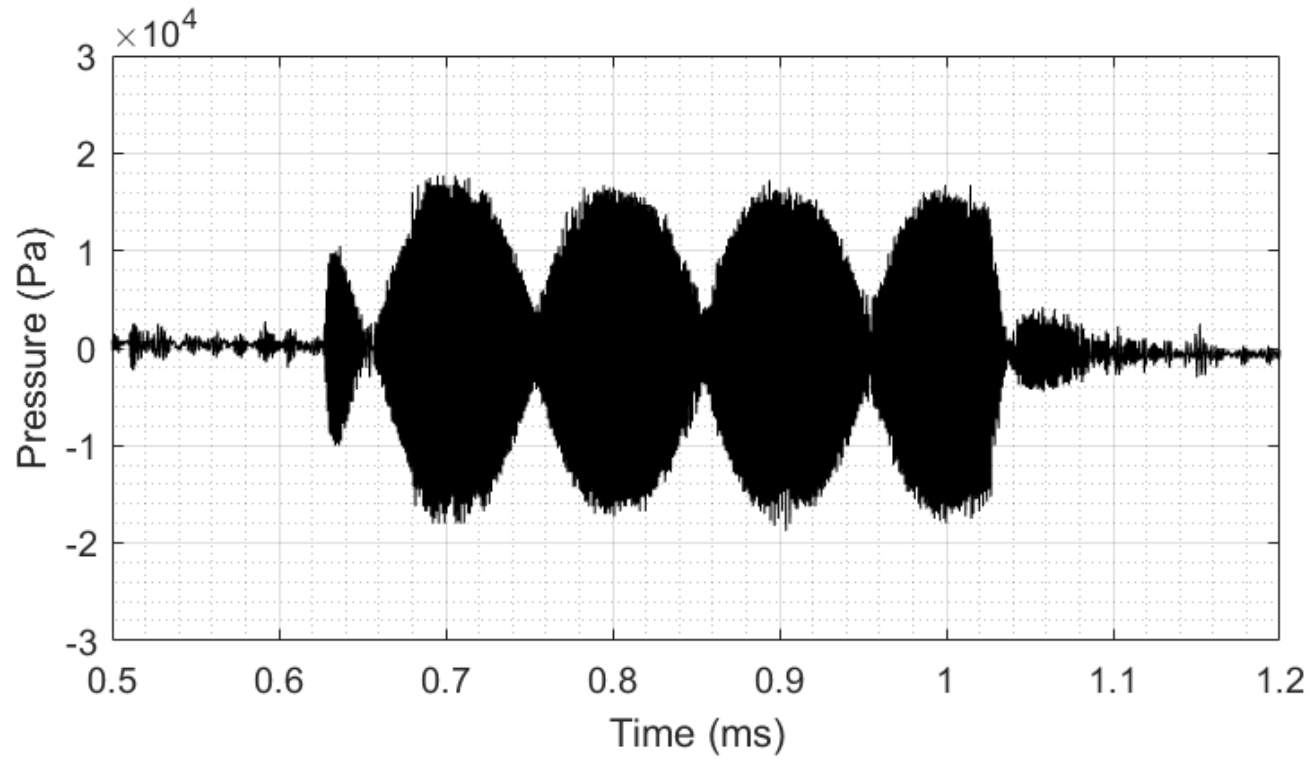

(a)

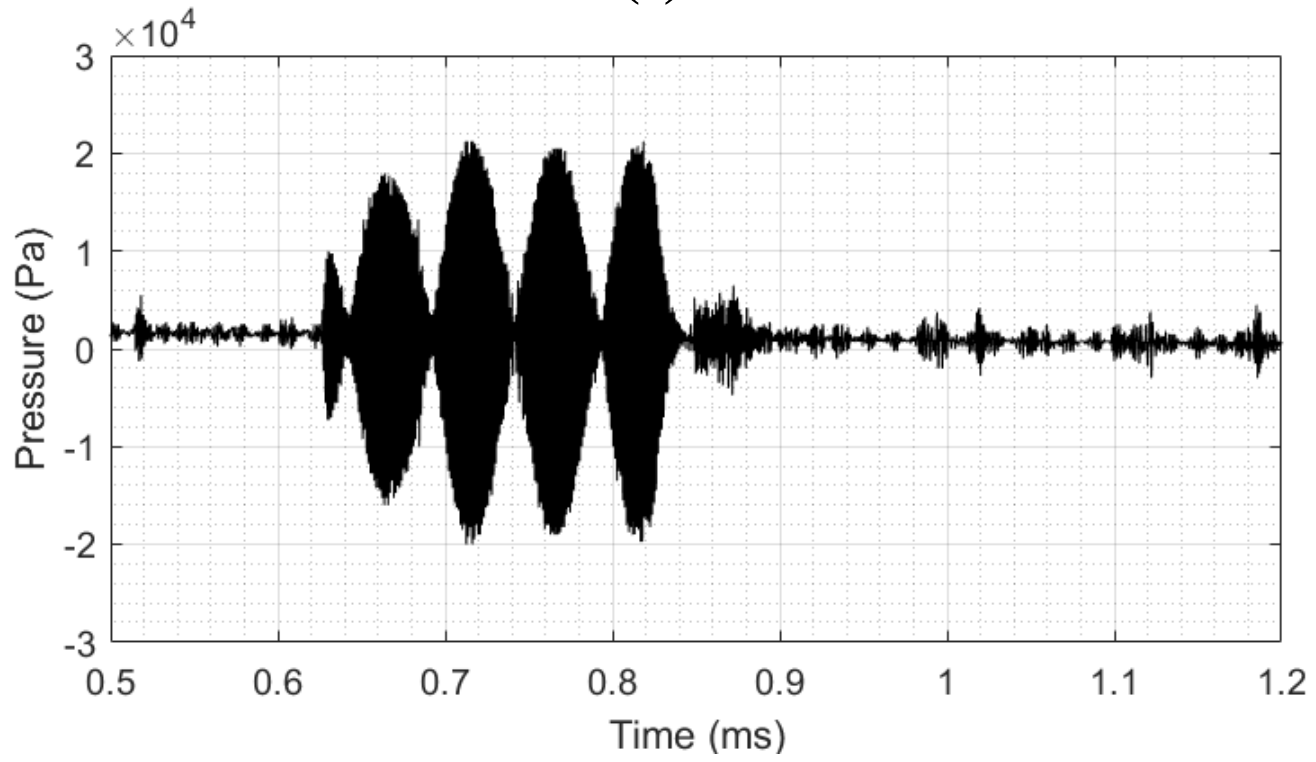

(b)

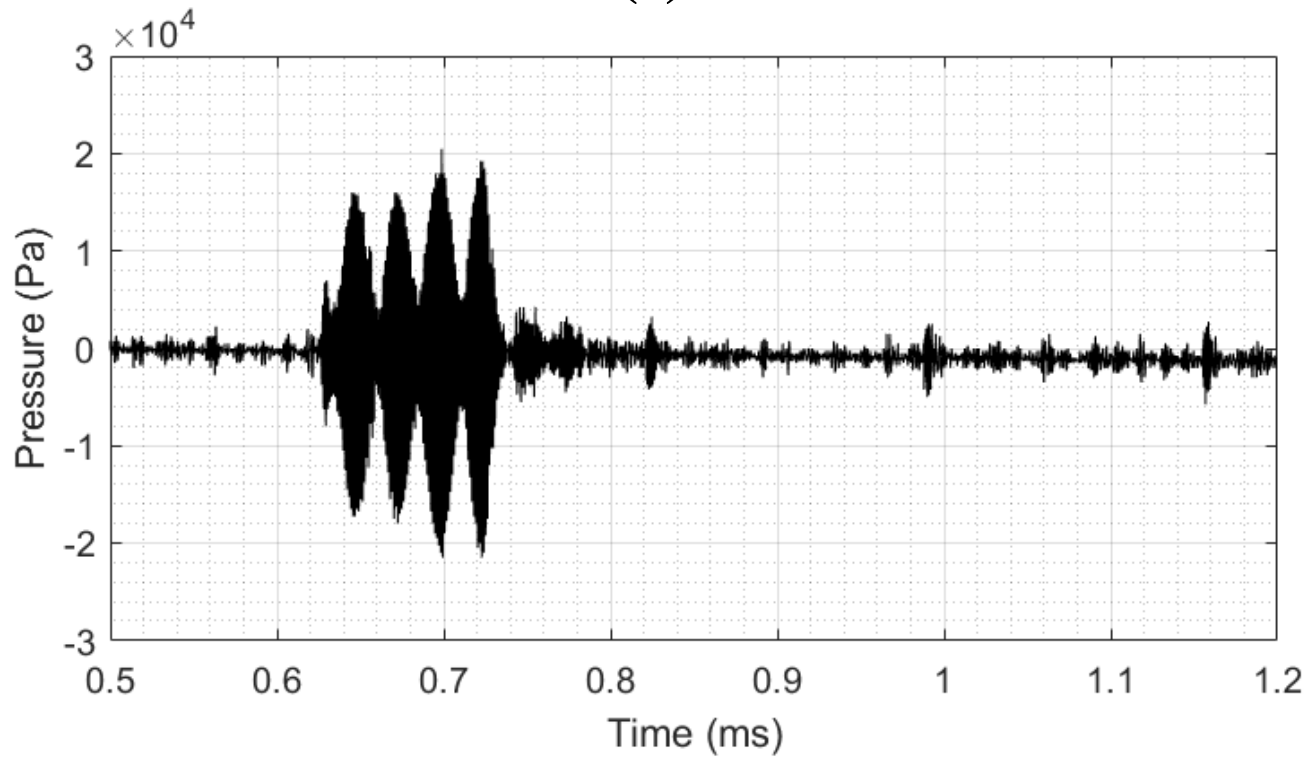

(c)

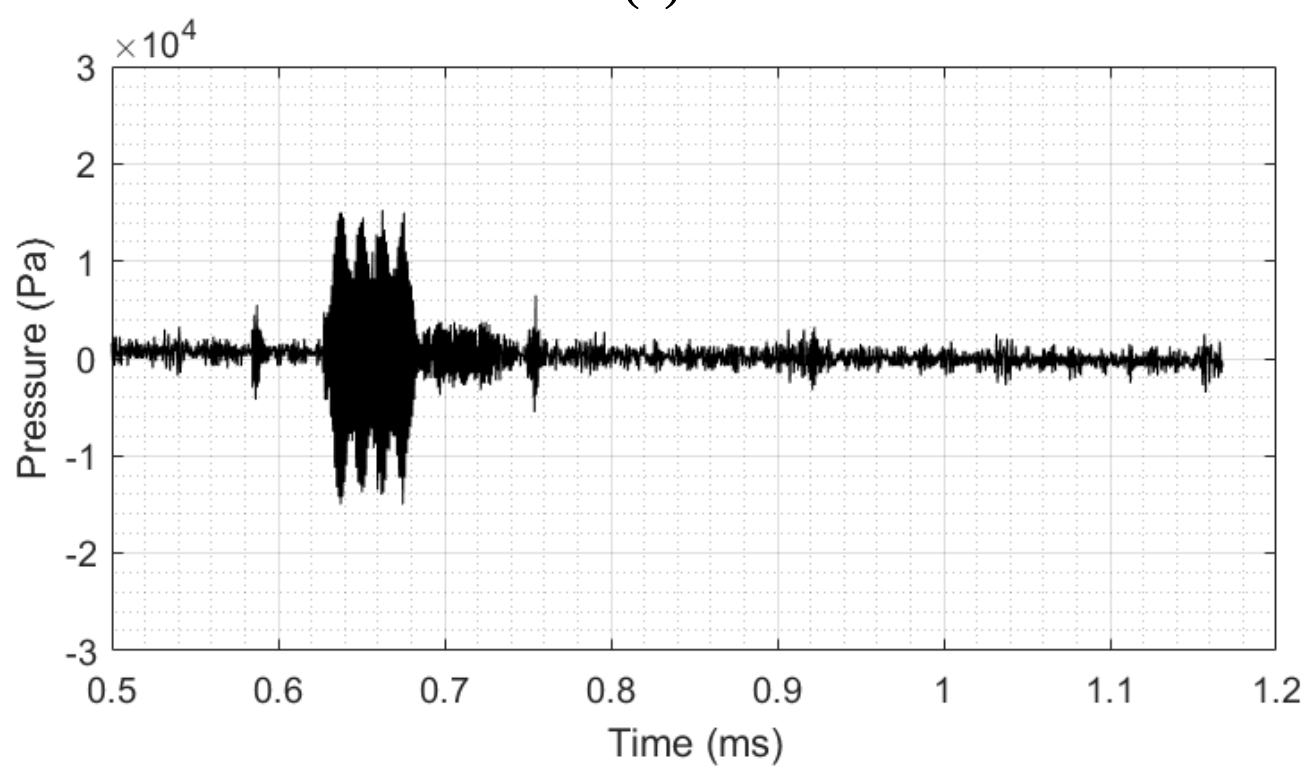

(d)

Fig. 9. The pressure waves, $p_N(t)$, sensed by the needle hydrophone at $1\ m$ away from the transducer, resulting from the transmission of $s(t)$. The subscript "N" represents the needle hydrophone. These waves are propagating waves in the primary acoustic field, which will result in 4-cycle pulses at (a) 10 kHz (b) 20 kHz (c) 40 kHz (d) 80 kHz difference frequencies in the secondary acoustic field after nonlinear interaction.

## *B. Secondary acoustic field measurements*

After we measured the pressure in the primary acoustic field, we removed the needle hydrophone from the water. Then, we measured the pressure at the difference frequency in the secondary acoustic field resulting from transmitting the signals representing the “+1” symbol shown in Fig. 1. We used the D/140/H hydrophone in these measurements.

Figures 10(a) and 10(b) show one of the pressure signals measured by the D/140/H hydrophone and its frequency spectrum, respectively, before any post-processing technique is applied. Figure 10 illustrates that the transmitted signal cannot find sufficient distance to travel for the higher frequency components to attenuate. The dominance of higher frequency components at and around the primary frequency of 855 kHz, is evident in Fig. 10(b). Additionally, the spectrum indicates that the filtering performed with the LNA is inadequate in removing these higher frequency components, highlighting the need for further filtering through post-processing operations.

We post-processed the pressure signals measured by the D/140/H hydrophone in MATLAB to remove the high-frequency components shown in Fig. 10(b). We applied low-pass and high-pass Butterworth filters, as detailed in Sec. IV. Figures 11(a)-11(c) show the resulting 2, 4, and 8-cycle pulses, $p_d(t)$, generated in secondary acoustic field. These signals are generated due to the transmissions shown in Figs. 1(c) and 1(d).

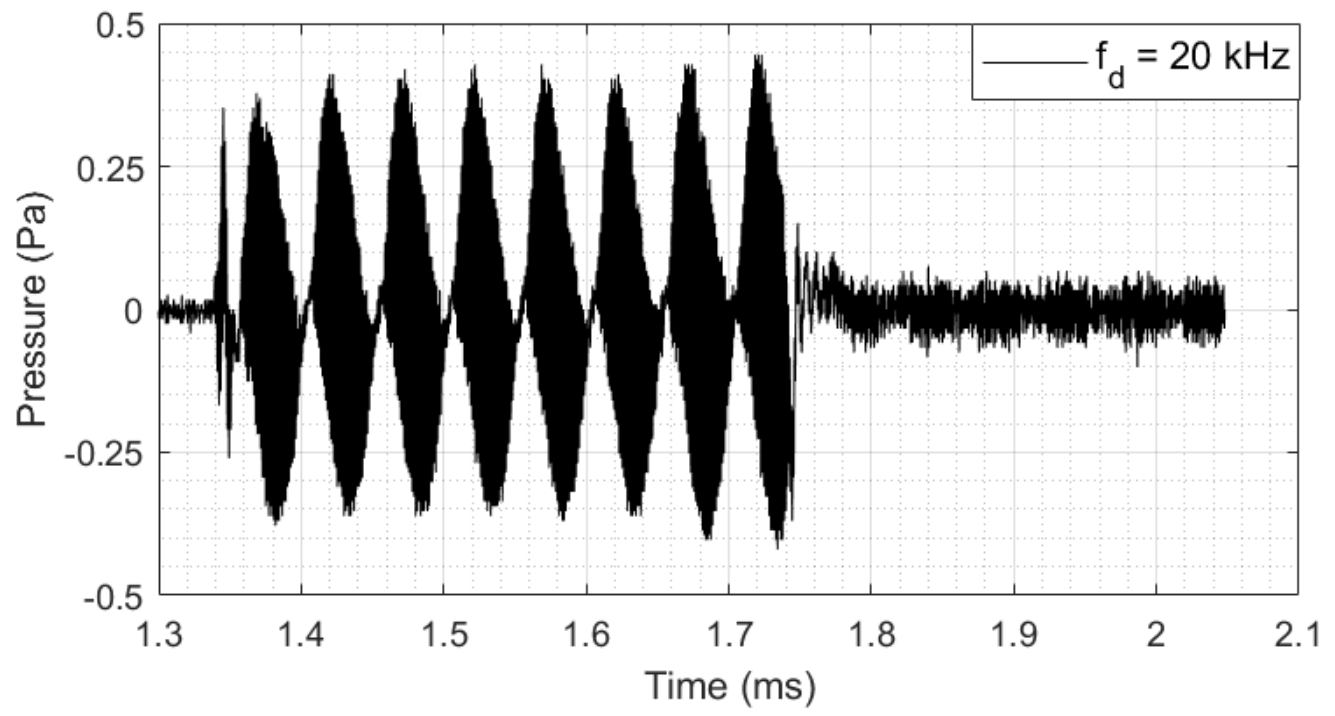

(a)

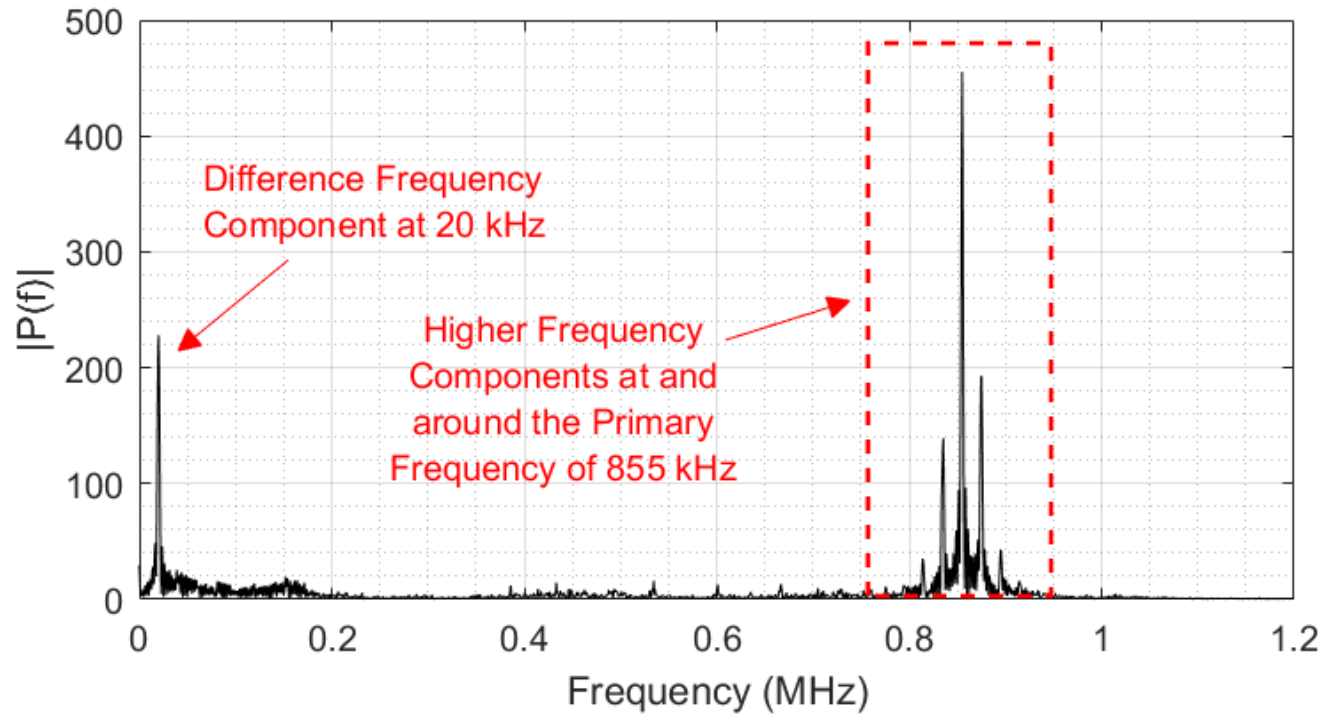

(b)

Fig. 10. (a) The pressure signal measured by the D/140/H hydrophone at 2 $m$ away from the transducer before postprocessing, resulting from the transmission of $s(t)$ to generate 8-cycle difference frequency signal at 20 kHz. (b) Its frequency spectrum. The difference frequency component at 20 kHz and higher frequency components at and around the primary frequency, 855 kHz, are shown.

We generated well-defined sinusoidal pulses, as shown in Fig. 11, in the secondary acoustic field, but with a slight increase in pulse durations. This increase is attributed to the combined effect of the filtering operation and the transducer transient response. We also measured that the frequency of these pulses is exactly 20 kHz, which is the difference frequency.

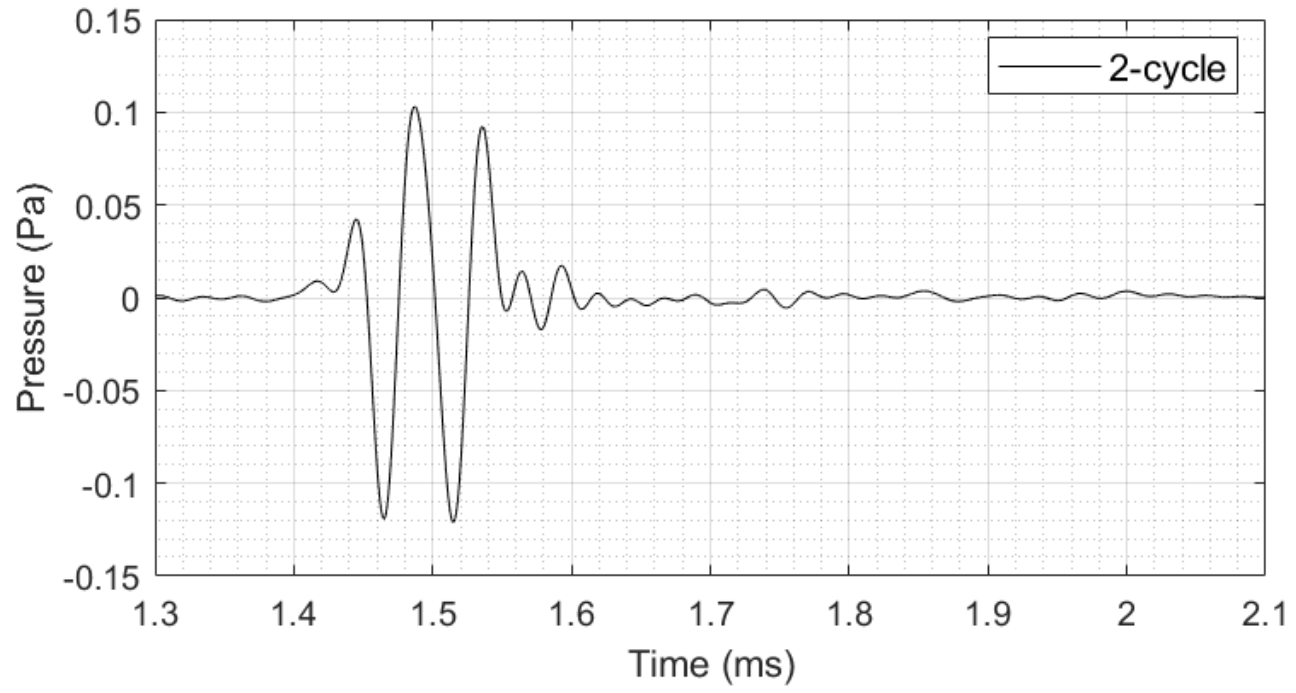

(a)

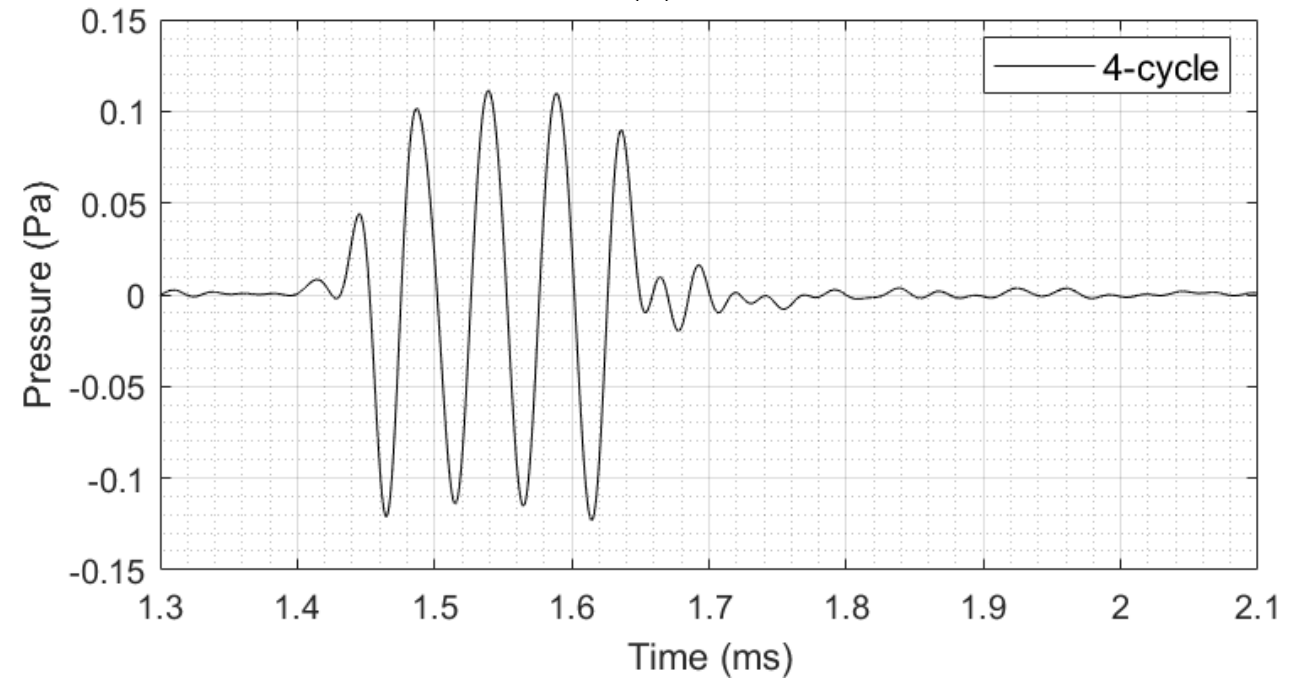

(b)

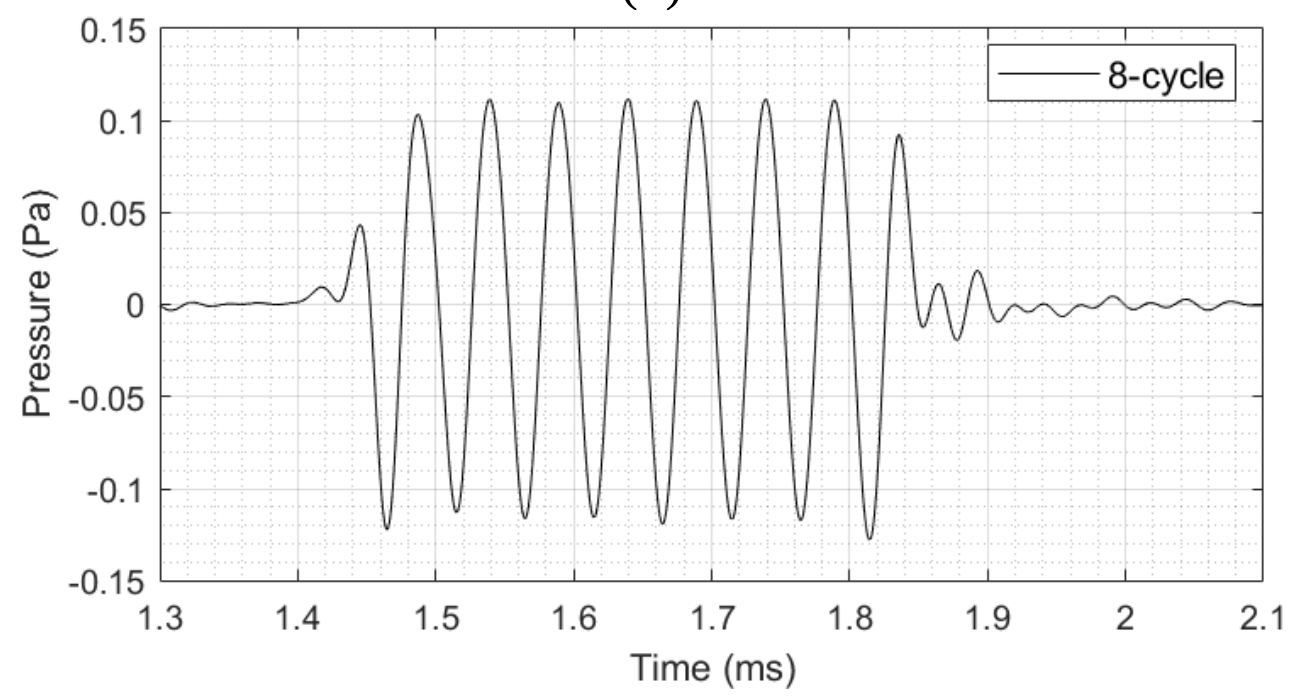

(c)

Fig. 11. The pressure pulses with (a) 2-cycle, (b) 4-cycle, and (c) 8-cycle at 20 kHz difference frequency, $p_d(t)$, measured by the D/140/H hydrophone at 2 $m$ away from the transducer, resulting from the transmission of $s(t)$ at different time instants. The subscript “d” represents the difference frequency.

Similarly, we measured the pressures at the difference frequency by the D/140/H hydrophone for the transmitted signals shown in Fig. 2, representing the “-1” symbol. Figures 12(a)-12(c) show the resulting 2, 4, and 8-cycle pulses, $p_d'(t)$, respectively. Upon comparison of the pulses in Fig. 11 and Fig. 12, we observed similar results.

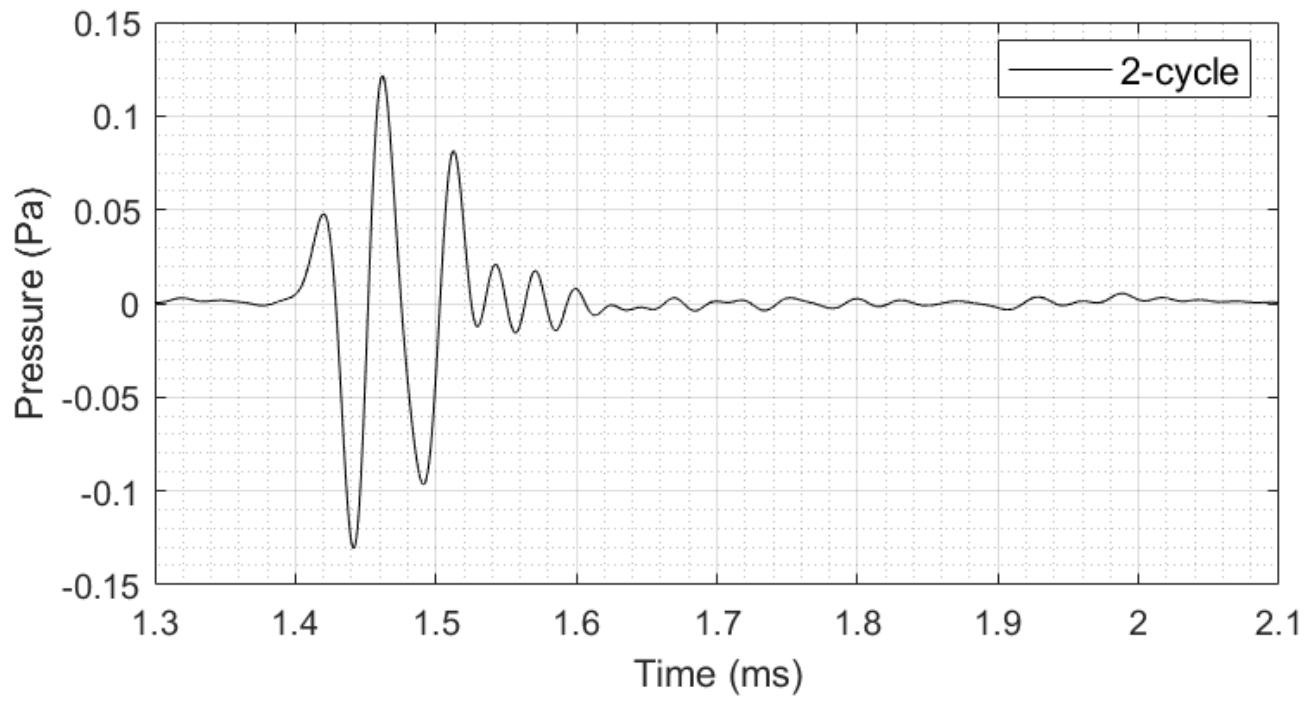

(a)

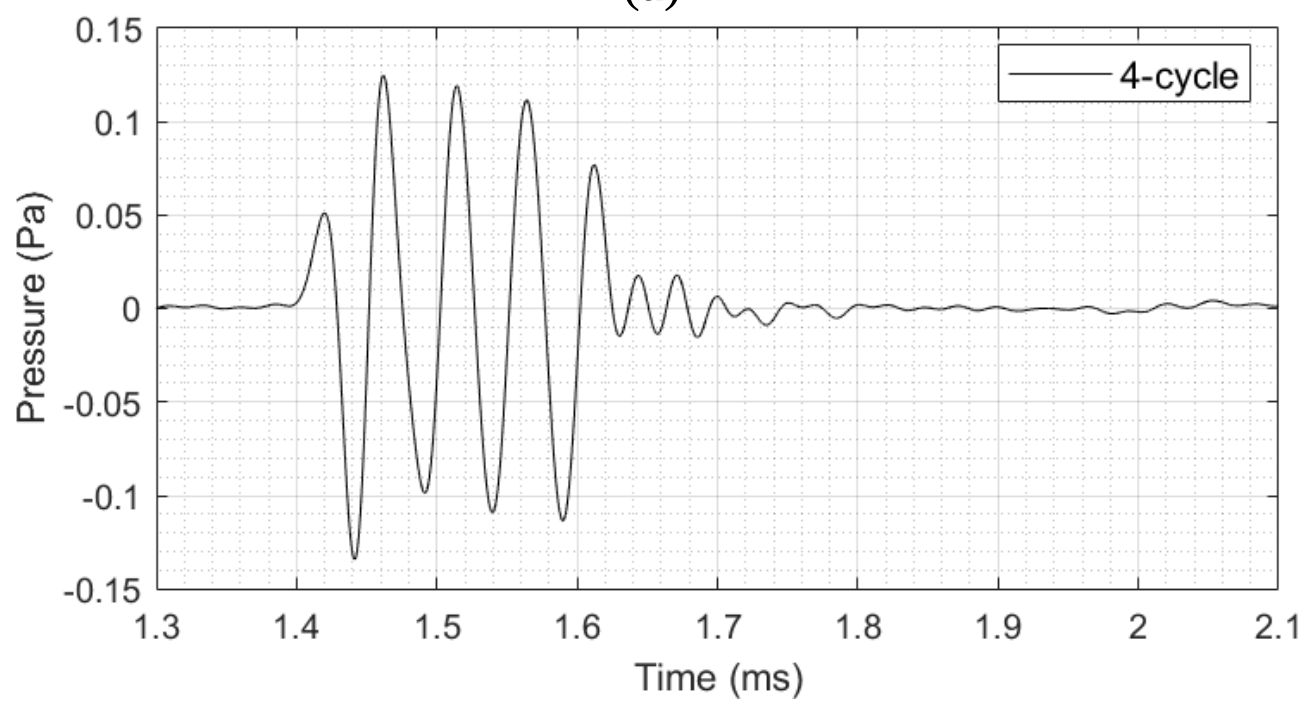

(b)

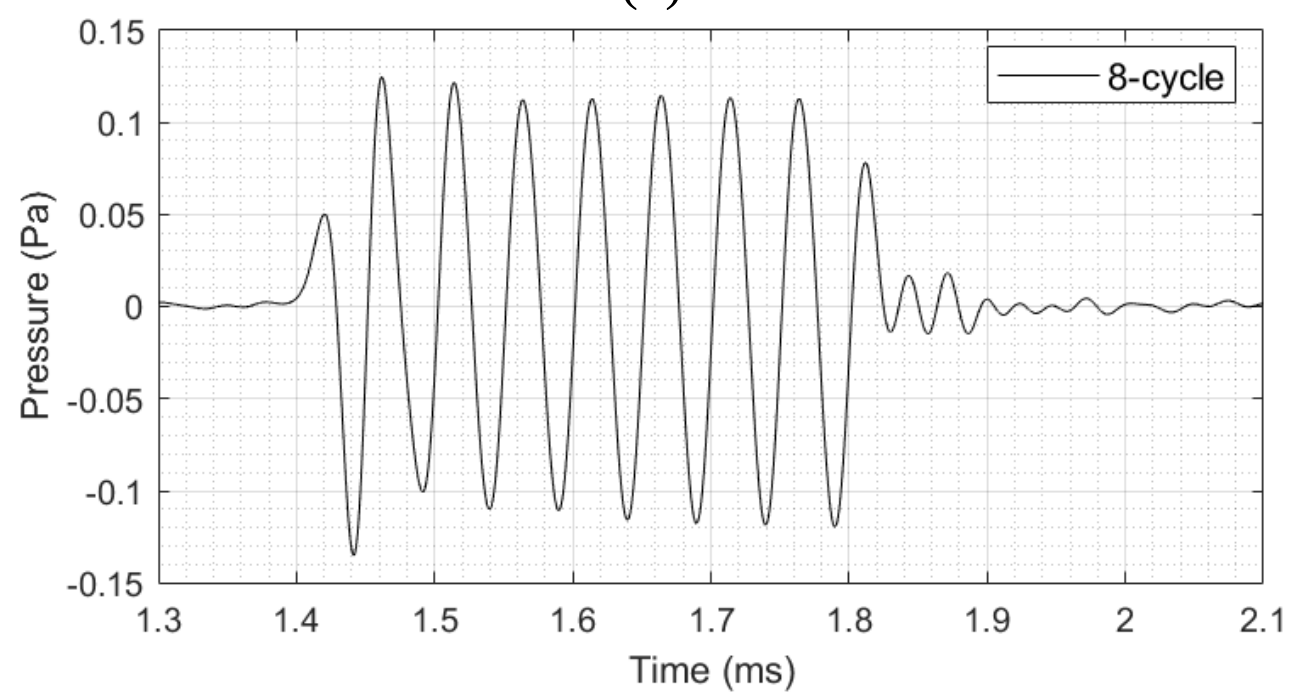

(c)

Fig. 12. The pressure pulses with (a) 2-cycle, (b) 4-cycle, and (c) 8-cycle at 20 kHz difference frequency, $p'_d(t)$, measured by the D/140/H hydrophone at $2\ m$ away from the transducer, resulting from the transmission of $s'(t)$.

The transmitted signals shown in Fig. 1 and Fig. 2, representing the "+1" and "-1" symbols, appear to generate similar pulses at the difference frequency. However, upon closer examination, a significant difference emerges: a 180° phase difference between these pulses at the difference frequency. This phase difference is delineated at two different time instants in Fig. 13, where the 8-cycle pulses at 20 kHz from Figs. 11(c) and 12(c) are depicted. This means that when the pressure reaches its maximum in a specific period of $p_d(t)$, it descends to its minimum within the same period of $p'_d(t)$ or vice versa. That is to say, these pulses are out-of-phase. We used this ability of producing two DFWs with a 180° phase difference for the coded DFW generation.

Furthermore, we measured that the peak pressure of $p_d(t)$ and $p'_d(t)$ is approximately 0.12 Pa at 20 kHz for all pulse lengths.

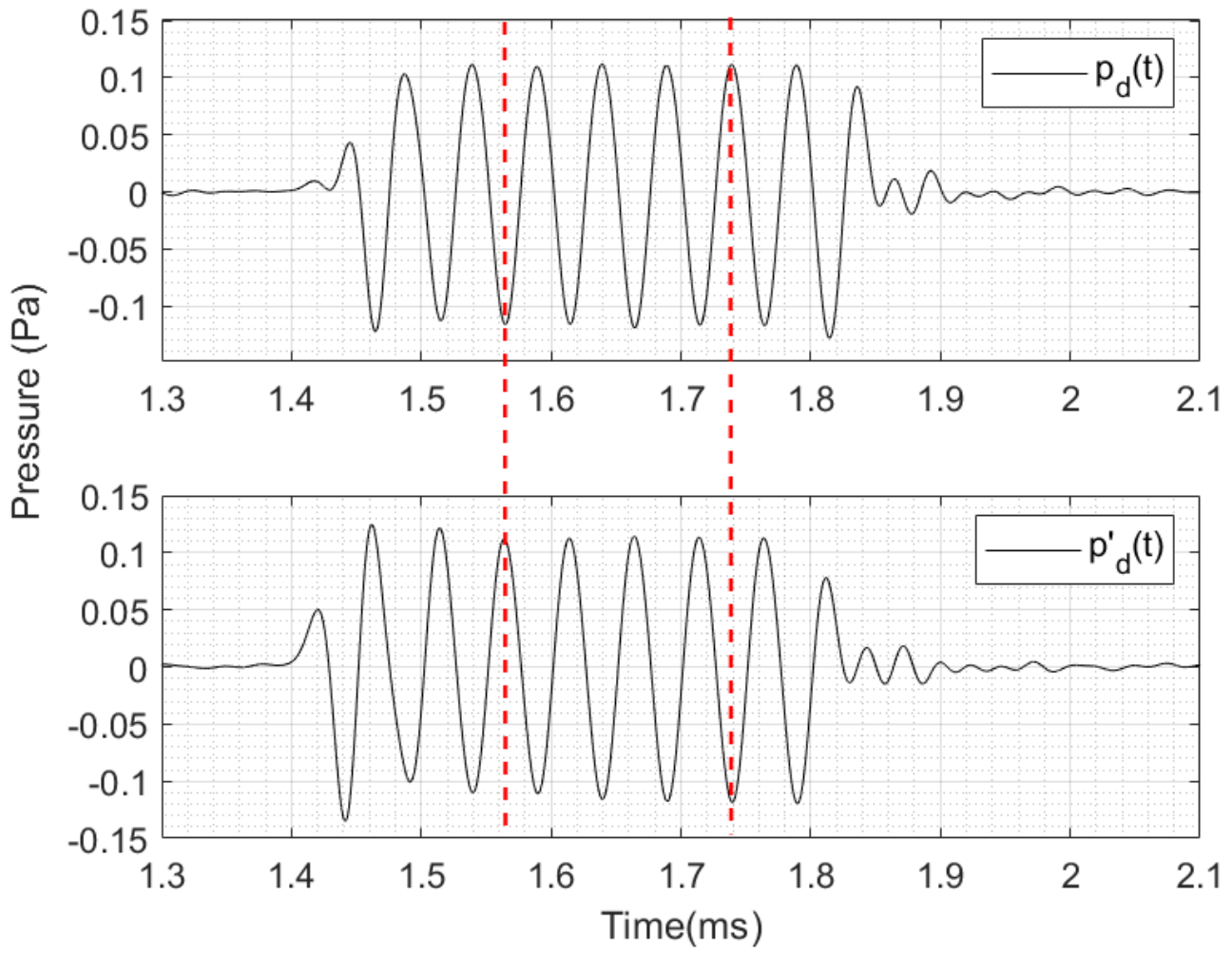

Fig. 13. The pressure pulses at 20 kHz difference frequency, $p_d(t)$ and $p'_d(t)$, measured by the D/140/H hydrophone at $2\ m$ away from the transducer, resulting from the transmission of (a) $s(t)$ and (b) $s'(t)$. The 180° phase difference is delineated at two different time instants.

### C. *Secondary source level measurements*

We also measured the source levels at the difference frequencies. We measured that the source levels are 92.6, 104.6, 112.6, and 114.7 dB at 10, 20, 40, and 80 kHz, respectively [52]. These measured values would be much higher if we performed the measurements in the far-field of the primary beam. Since we had to measure the DFW at the 2 m range due to the limited test tank size, the secondary wave generation in the parametric array is not fully exploited at this distance, resulting in lower source levels than what is expected.

### D. *Beamwidth measurements*

We measured the half-power beamwidth [7] of the parametric array. Initially, we removed the needle hydrophone from the test tank and proceeded with successive measurements to generate 8-cycle pulses at 20 kHz difference frequency. We moved the D/140/H hydrophone along the axis perpendicular to the propagation axis with 2.5 cm intervals at each measurement. At each position of the hydrophone, we measured the received pulse at 20 kHz and post-processed it as detailed in Sec. IV.

Then, we correlated the received pulses at difference frequency with reference signals. To generate a reference signal, we created a sinusoidal signal of the same cycle as the received pulse in MATLAB. Then, we applied the same post-processing techniques to this signal in the same way that these techniques are applied to the measured DFWs to include the processing effects. We used the resulting signals as the reference signals in our study. We calculated the root mean square (rms) pressure at this position. We repeated this process for each hydrophone position. We normalized the rms pressure values with respect to that measured at the median hydrophone position, where the rms pressure is maximum.

Finally, we measured the beamwidth of the parametric array. Figure 14 shows the measured beamwidth and a -3 dB level to illustrate the half-power points. We measured the beamwidth of the parametric array as 7.5° as shown in Fig. 14. This result demonstrates that the measured half-power beamwidth, 7.5°, is very close to the theoretically calculated beamwidth, 4° [7]. There are no sidelobes in the measured beamwidth at the difference frequency.

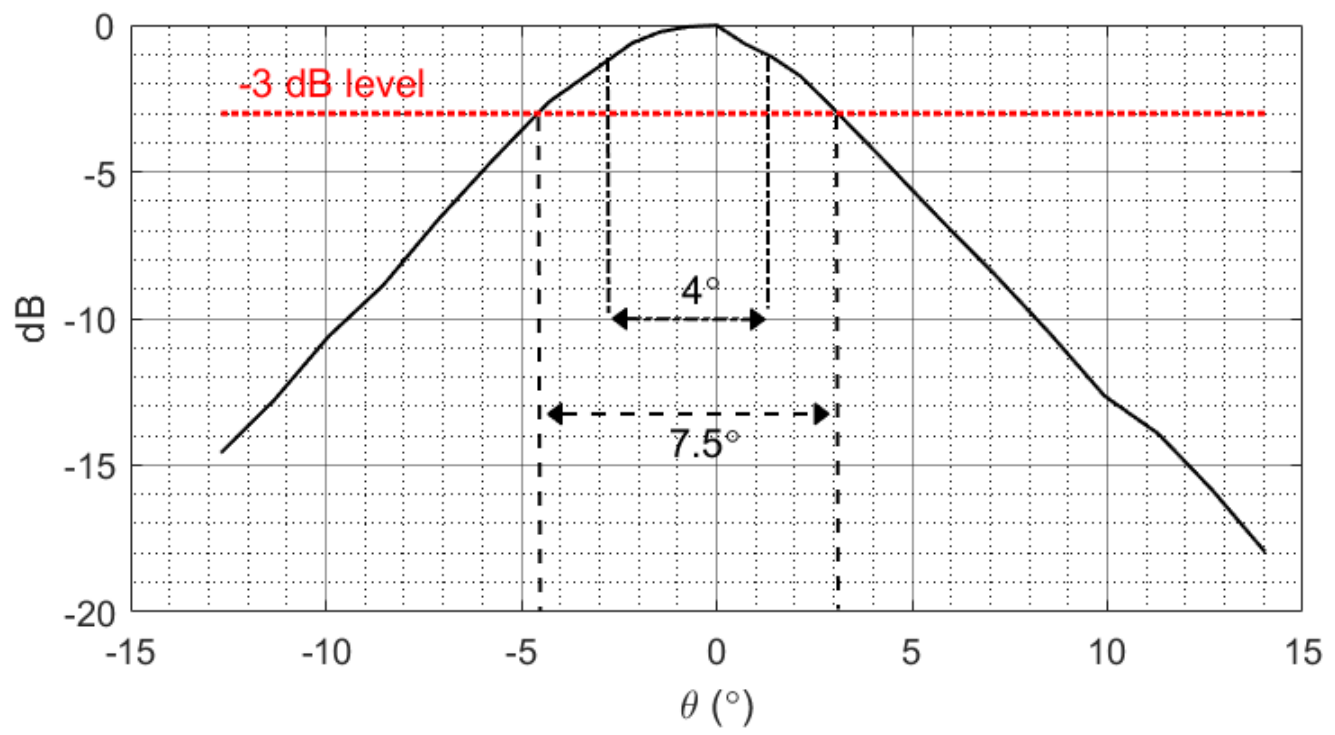

Fig. 14. The measured beamwidth of the parametric array used in this study. The beamwidth, 7.5°, was measured through the processing of the received pressure pulses at 20 kHz difference frequency. The D/140/H hydrophone was used in the beamwidth measurements. A -3 dB level is plotted to illustrate the half-power points. The theoretically calculated beamwidth, 4°, is also shown.

## VI. Discussion Of Results And Conclusion

This study employs a parametric array and introduces the wideband pulse generation at the difference frequency for underwater acoustic applications. We measured well-defined 2, 4, and 8-cycle pulses at 10, 20, 40, and 80 kHz, while pulses with wider bandwidths (0.5 and 1-cycle pulses) were measured with an increased distortion. We also showed that phase-sensitive generation of bursts in the secondary acoustic field is possible.

### A. *Secondary acoustic field signals*

The experimental results revealed an increase in the pressure of the DFW with increasing difference frequency from 10 kHz to 80 kHz. However, contrary to the prediction of Eq. (2) suggesting a fourfold increase in pressure as the difference frequency doubles, the experimental results did not exhibit such a trend, as depicted in Fig. 15.

There are several reasons for the discrepancy between the measured and predicted pressure amplitudes. First, the measurement distance, 2 m, is significantly small to satisfy the approximation made in deriving Eq. (2), in which the nonlinear interaction is assumed to be confined to the near field of the primary beam [13]. In this study, we could not fully exploit the nonlinearity within the near field. The other reason is that the Eq. (2) predicts the DFW pressure when the nonlinear interaction region, characterized by $R_a$, is smaller than the $R_f$ [13]. However, $R_a$ is much greater than $R_f$ in our study. The other reason is that Eq. (2) is derived under the assumption of transducers with infinite bandwidth, whereas the transducer used in this study has finite bandwidth. The last reason can be attributed to the removal of harmonic components in the measured secondary acoustic field using a lowpass filter with a cutoff frequency twice the difference frequency. Consequently, the measured pressure amplitude differs from the theoretically predicted outcome. This effect aggravates for difference frequencies larger than 40 kHz, where the increase in DFW pressure remains limited as shown in Figs. 15(f) and 15(g).

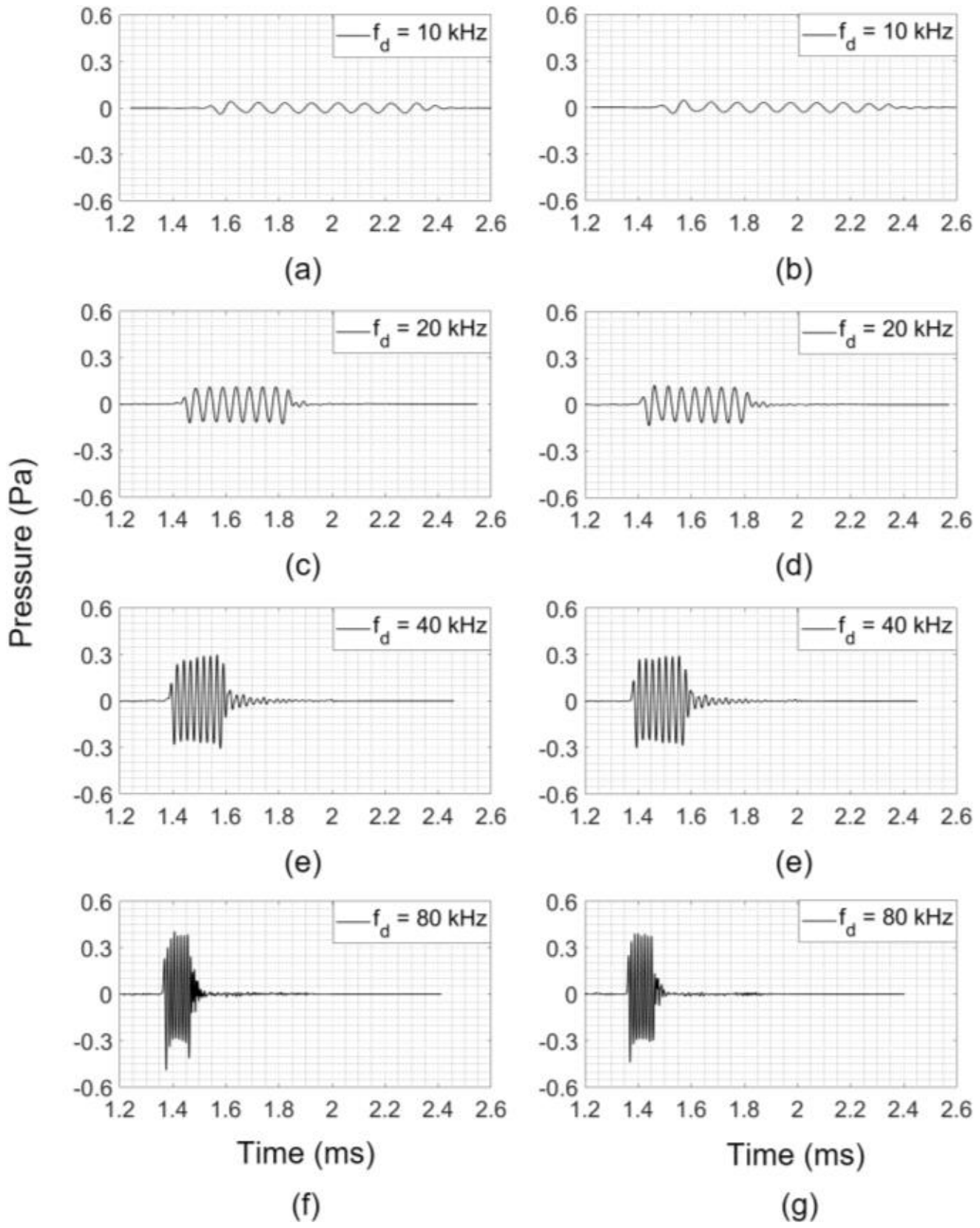

Fig. 15. The pressure pulsed waves at various difference frequencies sensed by the D/140/H hydrophone. (a, c, e, g) $p_d(t)$ and (b, d, f, h) $p'_d(t)$ at 10, 20, 40, and 80 kHz, respectively. $p_d(t)$ and $p'_d(t)$ are out-of-phase signals, which have 180° phase difference.

### B. *The signal quality*

We assess the quality of the signals measured at the difference frequency in the secondary acoustic field. For this purpose, we investigated the normalized cross-correlation between the measured and reference signals, which is a measure of the similarity between these signals. We obtained the reference signals as described in Sec. V. D.

Figure 16 illustrates a 2-cycle pulse measured at 40 kHz in the secondary acoustic field, its reference signal, and their correlation, respectively. The peak of the normalized correlation output, shown in the dotted circle in Fig. 16, for the 2-cycle pulse at 40 kHz is 0.92, indicating a high signal quality. It occurs when $\tau$ is 1.35 ms, where $\tau$ represents the time delay. The time delay axis was adjusted such that the delay corresponding to the peak shows the arrival time of the received signal at the hydrophone. The time at which the peak correlation is observed agrees with the arrival time of the measured signal, as seen in Fig. 16.

We observed that the correlation improves as the pulse duration increases. The peak of the normalized correlation outputs for the 1, 2, 4, and 8-cycle pulses at 40 kHz are 0.89, 0.92, 0.94, and 0.97, respectively. Additionally, the correlation output remains almost the same at different frequencies for the same pulse duration.

It is also expected that generating the reference signals using the measured data from the near field instead of using MATLAB may improve their similarity to the received signals, and consequently, further increase the correlation output [3].

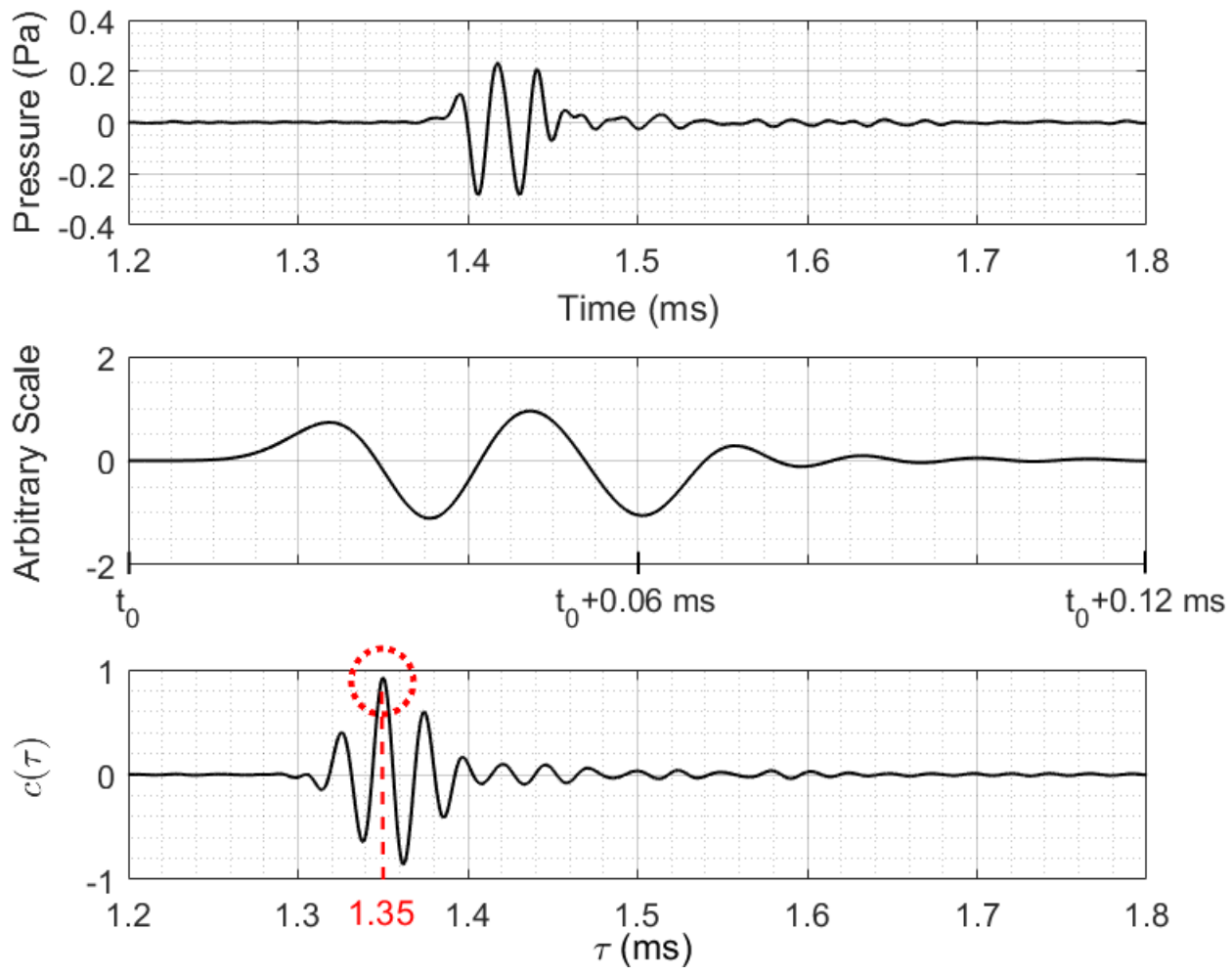

Fig. 16. The pressure signal at 40 kHz measured by the D/140/H hydrophone were correlated with the corresponding reference signal. (a) Measured 2-cycle pulse (b) Its reference signal, which is generated in MATLAB. The post-processing techniques are applied to the reference signal in the same way that they are applied to the measured pulse. The reference signal duration is 0.12 ms and each division corresponds to 0.02 ms. (c) Normalized correlation output with respect to time delay. The peak output and corresponding time delay are shown within a dotted circle and with a dashed line, respectively.

### C. *Limitations and potential improvements*

The measured DFW contains primary frequency components despite using LNA as a bandpass filter during the measurements. Because, the primary field is significantly high at the measurement range of 2 m. We implemented software low-pass and high-pass filters to obtain well-defined DFWs. The environments where primary frequency signals can propagate sufficiently may lead to complete attenuation of the primary components in the secondary acoustic field. This, in turn, will eliminate the need for additional post-processing operations. This issue must be investigated through seagoing measurements for possible improvements.

The transmitted signals aimed at generating 0.5 and 1-cycle pulses in the secondary acoustic field fails to attain their peak values, which causes significant distortion in the secondary acoustic field. Various modulation techniques have been employed as pre-processing methods in the literature to reduce distortion in the DFWs. In our study, we utilized the SQRAM technique. Our measurement results show that there is room for further improvement in reducing signal distortion. In this context, determination and implementation of the modulation technique represent a potential area for distortion reduction.

We observed that there is 180° phase difference between the measured wideband pulses at the difference frequency representing “+1” and “-1” symbols. This phase difference provides an insight into the potential for generating binary phase shift keying modulated coded signals in the secondary acoustic field.

## Acknowledgments

This work was supported by the Scientific and Technological Research Council of Turkey (TUBITAK) under Grant No. 119E509. The authors would like to thank Professor Ekmel Özbay and his research group at NANOTAM, Bilkent University for providing valuable resources during the measurements.